\documentclass[10pt, journal, letterpaper, twocolum, twoside, compsoc]{IEEEtran}
\usepackage{graphicx}
\usepackage{balance}  % for  \balance command ON LAST PAGE  (only there!)
\usepackage[ruled,vlined]{algorithm2e}
\usepackage{algorithmicx}
\usepackage{algpseudocode}
\usepackage{amsmath,amssymb,amsfonts,bm}
\usepackage{multirow}
\usepackage{upgreek}
\SetCommentSty{mycommfont}
\usepackage{times}
\usepackage{subfigure}
\usepackage{array}
\usepackage{colortbl}
\usepackage[table,xcdraw]{xcolor}

\definecolor{mygray}{gray}{.5}
\definecolor{mypink}{rgb}{.99,.91,.95}
\definecolor{mycyan}{cmyk}{.3,0,0,0}

\begin{document}

% ****************** TITLE ****************************************

\title{A Hybrid Data Cleaning Framework using Markov Logic Networks\vspace*{-0.1in}}

\author{Yunjun~Gao,%~\IEEEmembership{Member,~IEEE,}
        Congcong~Ge,
        Xiaoye~Miao,
        Haobo~Wang,
        Bin~Yao,%~\IEEEmembership{Member,~IEEE,}
        Qing~Li

\IEEEcompsocitemizethanks{
\IEEEcompsocthanksitem Y. Gao, C. Ge, and H. Wang are with the College of Computer Science, Zhejiang University, Hangzhou 310027, China, E-mail:\{gaoyj, gcc wanghaobo\}@zju.edu.cn.
\IEEEcompsocthanksitem X. Miao is with the Center for Data Science, Zhejiang University, Hangzhou 310027, China, E-mail: miaoxy@zju.edu.cn.
\IEEEcompsocthanksitem B. Yao is with the Department of Computer Science and Engineering, Shanghai Jiao Tong University, Shanghai 200000, China, E-mail: yaobin@cs.sjtu.edu.cn.
\IEEEcompsocthanksitem Q. Li is with the Department of Computer Science, City University of Hong Kong, Hong Kong, China, E-mail: itqli@cityu.edu.hk.
}
%\thanks{Manuscript received xxxx, xxxx; revised xxxx, xxxx.}
}

%\markboth{IEEE Transactions on Knowledge and Data Engineering,~Vol.~XX, No.~XX, XXX~XXXX}{Gao \MakeLowercase{\textit{et al.}}: A Hybrid Data Cleaning Framework using Markov Logic Networks with Multiple Data Versions}

\IEEEtitleabstractindextext{
\begin{abstract}
With the increase of dirty data, data cleaning turns into a crux of data analysis. Most of the existing algorithms rely on either qualitative techniques (e.g., data rules) or quantitative ones (e.g., statistical methods).
%have many drawbacks, for example they have limited scalability to large-scale data, or they rely on external clean data information.
In this paper, we present a novel hybrid data cleaning framework on top of Markov logic networks (MLNs), termed as \textsf{MLNClean},  which is capable of cleaning both schema-level and instance-level errors.
\textsf{MLNClean} mainly consists of two cleaning stages, namely, first cleaning multiple data versions separately (each of which corresponds to one data rule), and then deriving the final clean data based on multiple data versions. Moreover, we propose a series of techniques/concepts, e.g., the MLN index, the concepts of \emph{reliability score} and \emph{fusion score}, to facilitate the cleaning process.
Extensive experimental results on both real and synthetic datasets demonstrate the superiority of \textsf{MLNClean} to the state-of-the-art approach in terms of both accuracy and efficiency.
\end{abstract}

\begin{IEEEkeywords}
data cleaning, MLN, qualitative, quantitative
\end{IEEEkeywords}}

\maketitle

\newtheorem{definition}{Definition}
\newtheorem{example}{Example}

\IEEEdisplaynontitleabstractindextext

\IEEEpeerreviewmaketitle

%\author{Yunjun Gao$^{\dagger}$$^{\sharp}$, Congcong Ge$^{\dagger}$, Xiaoye Miao$^{\ddagger}$$^{*}$, Haobo Wang$^{\dagger}$, Bin Yao$^{\mathsection}$, Qing Li$^{*}$ \\
%\and
%\alignauthor
%    \affaddr{{\large$^{\dagger}$}College of Computer Science, Zhejiang University, Hangzhou, China}\\
%    \affaddr{{\large$^{\ddagger}$}Center for Data Science, Zhejiang University, Hangzhou, China}\\
%    \affaddr{{\large$^{\sharp}$}Alibaba--Zhejiang University Joint Institute of Frontier Technologies, Hangzhou, China}\\
%    \affaddr{{\large$^{*}$}Department of Computer Science, City University of Hong Kong, Hong Kong, China}\\
%    \affaddr{
%        {\large$^{\mathsection}$}Department of Computer Science and Engineering, Shanghai Jiao Tong University, Shanghai, China\\
%        \email{
%            {\large$^{\dagger}$}{\large$^{\ddagger}$}\{gaoyj, gcc, miaoxy, wanghaobo\}@zju.edu.cn~
%            {\large$^{\mathsection}$}yaobin@cs.sjtu.edu.cn~
%            {\large$^{*}$}itqli@cityu.edu.hk
%        }
%    }
%}

\section{Introduction}

Data analysis benefits from a wide variety of reliable information.
The existence of dirty data not only leads to erroneous decisions or unreliable analysis but probably causes a blow to the corporate economy~\cite{eckerson2002data}.
As a consequence, there has been a surge of interest from both industry and academia on data cleaning~\cite{chu2016data}.
The purpose of data cleaning is generally to correct errors, to remove duplicate information, and to provide data consistency. It usually includes two steps, i.e.,  error detecting and error repairing.
The first step is to find where dirty data hide, and the second one is to correct dirty data detected in the previous step.
%Data cleaning is the process of re-examining and verifying data, which purpose is to remove duplicate information, correct errors, and provide data consistency. Data cleaning usually consists of two stages: (i) error detecting and (ii) error repairing. The goal of error detecting is to find where dirty data hidden and the goal of error repairing is to correct dirty data detected in the previous stage. Basically, data cleaning can be divided into two major categories: (i) qualitative techniques and (ii) quantitative techniques.

\begin{example}
\label{example:hai}
Table 1 depicts a group of sampled tuples from a dirty hospital information dataset $T$. It contains four attributes, including hospital name $($HN$)$, city $($CT$)$, state $($ST$)$, and phone number $($PN$)$. The dataset needs to comply with three integrity constraints, i.e., one functional dependency $($FD$)$, one denial constraint $($DC$)$, and one conditional functional dependency $($CFD$)$.
{\small
\vspace*{0.1in}
\noindent
\\\emph{($r_1$) FD:  CT $\Rightarrow$  ST
\\($r_2$) DC: $\forall t, t' \in T, \lnot$(PN($t.v$) = PN($t'.v$) $\land$ ST($t.v$) $\neq$ ST($t'.v$))
\\($r_3$) CFD: HN(``ELIZA"), CT(``BOAZ")$\Rightarrow$PN(``2567688400")}

\vspace*{-0.1in}
\noindent
}

Specifically, the rule $r_1$ means that a city uniquely determines a state, the rule $r_2$ indicates that two hospitals located in different states have different phone numbers, and the rule $r_3$ means that a hospital named ``ELIZA" and located in city ``BOAZ", has a specific phone number ``2567688400". Errors appeared in the tuples are highlighted in colored cells, and they can be treated from two different levels, i.e., the schema-level and instance-level $\cite{rahm2000data}$.

The schema-level errors refer to the values that violate integrity constraints.
For example, tuples $t_4$, $t_5$ and $t_6$ are violated on the attribute ST w.r.t. $r_1$.
The instance level errors contain \emph{replacement errors}, \emph{typos}, and \emph{duplicates} in this example.
In particular, the replacement error signifies that a value is incorrectly recorded as another value. That is, the value is completely wrong.
%This error may occur when users select a wrong value from an attribute domain which contains multiple values.
For instance, $t_3$.$[$CT$]$ being ``DOTHAN'' is a replacement error, the correct value should be ``BOAZ'' in this cell.
Typos, also called misprints, are caused by the typing process. For example, $t_2$.$[$CT$]$ being ``DOTH'' is a typo, and the correct value should be ``DOTHAN'' in this cell. In addition, duplicates indicate that there are multiple tuples corresponding to the same real entity, e.g., tuples $t_4$, $t_5$, and $t_6$.
\end{example}
%Specifically, the value ``DOTH" should be ``DOTHAN" on attribute CT of $t_2$, the value ``DOTHAN" should be ``BOAZ" on attribute CT of $t_3$, the value ``2565938310" should be ``2567688400" on attribute PN of $t_3$, and the value ``AK" should be ``AL" on attribute ST of $t_4$.

%\begin{table}[t]
%\centering
%\caption{A sample of hospital information dataset}
%\begin{tabular}{c|c|c|c|c|}\cline{2-5}\rowcolor{black}
%\multicolumn{1}{l|}{\textbf{}} & \color{white}\textbf{HN} & \color{white}\textbf{CT} & \color{white}\textbf{ST} & \color{white}\textbf{PN}\\ \cline{2-5}
%$t_1$   &ALABAMA&DOTHAN&AL&3347938701\\ \cline{2-5}
%$t_2$   &ALABAMA&\cellcolor{mypink}DOTH&AL&3347938701\\ \cline{2-5}
%$t_3$   &ELIZA&\cellcolor{mypink}DOTHAN&AL&\cellcolor{mypink}2565938310\\ \cline{2-5}
%$t_4$   &ELIZA&BOAZ&\cellcolor{mypink}AK&2567688400\\ \cline{2-5}  %AK is error，应为AL%
%$t_5$   &ELIZA&BOAZ&AL&2567688400\\ \cline{2-5}        %AK is error，应为AL%
%$t_6$   &ELIZA&BOAZ&AL&2567688400\\ \cline{2-5}
%\end{tabular}
%\label{table:intro_sample}
%\vspace*{-7mm}
%\end{table}

\begin{table}[t]
\centering
\small
\label{table:intro_sample}
\setlength{\tabcolsep}{7pt}
\caption{A Sample of a Hospital Information Dataset}
\hspace*{1mm}
\begin{tabular}{|r|c|c|c|c|} \hline
   \textbf{TID} & \textbf{HN} &  \textbf{CT} &  \textbf{ST} &   \textbf{PN}\\ \hline
$t_1$ & ALABAMA                                           & DOTHAN                                            & AL                                                & 3347938701                                        \\ \hline
$t_2$ & ALABAMA                                           & \cellcolor[HTML]{FFCCC9}DOTH                      & AL                                                & 3347938701                                        \\ \hline
$t_3$ & ELIZA                                             & \cellcolor[HTML]{FFCCC9}DOTHAN                    & AL                                                & \cellcolor[HTML]{FFCCC9}2567638410                \\ \hline
$t_4$ & ELIZA                                             & BOAZ                                              & \cellcolor[HTML]{FFCCC9}AK                        & 2567688400                                        \\ \hline
$t_5$ & ELIZA                                             & BOAZ                                              & AL                                                & 2567688400                                        \\ \hline
$t_6$ & ELIZA                                             & BOAZ                                              & AL                                                & 2567688400                                        \\ \hline
\end{tabular}
%\vspace*{-2mm}
\end{table}

Data cleaning methods can be divided into two major categories including qualitative techniques and quantitative ones. The qualitative techniques \cite{abedjan2015temporal, bertossi2005complexity, beskales2010sampling, beskales2013relative, bohannon2007conditional, bohannon2005cost, chu2013holistic, cong2007improving, dallachiesa2013nadeef, fan2008conditional, geerts2013llunatic, khayyat2015bigdansing,kolahi2009approximating, lopatenko2007efficient}
mainly rely on integrity constraints to express data quality rules.
They detect errors which violate integrity constraints, and repair errors with the principle of \emph{minimality} (i.e., minimizing the impact on the dataset by trying to preserve as many tuples as possible).
Take the dataset in Table \ref{table:intro_sample} as an example. Tuples $t_4$ and $t_5$ are violated on the attribute ST w.r.t. $r_1$.
Thus, according to the principle of minimality, it replaces the value ``AK" with ``AL" on the attribute ST of $t_4$, whereas it fails to repair the attributes CT and PN of $t_3$.
In addition, the attribute CT of $t_2$ cannot be repaired since it does not violate any rule.
In contrast, the second category containing quantitative techniques \cite{krishnan2016activeclean, mayfield2010eracer, yakout2013don} employs statistical approaches to detect possible errors, and finds probable repair candidates of errors based on the probability theory.
%The repair results meet statistical distribution.
Qualitative techniques guarantee that the cleaning results are in accordance with data quality rules, and quantitative techniques ensure that the cleaning results conform to statistical characteristics.
Recently, new attempts are made by \cite{prokoshynaSCMS15,rekatsinas2017holoclean}, which combine qualitative and quantitative techniques.
Nevertheless, they either consider only one kind of integrity constraints (e.g., FDs), or keep the error detecting and repairing steps in isolation and focus on data repairing process, incurring the redundant computation.
%But its limitation lies that, it only combines FDs and statistical methods without considering other types of ICs.
%Another work HoloClean \cite{rekatsinas2017holoclean} designs a compiler that unifies several data repair signals including ICs and \emph{external} data, and generates data repair plans based on the probability values of each error value, in other words, it repairs errors once at a time.
%Besides, HoloClean focuses on repairing stage but ignores detecting stage, it calls the existing approaches to execute detection instead. Hence, two stages of data cleaning are executed in isolation without taking into account the subsequent computation in the cleaning pipeline, incurring redundant computation. Besides, data repairing relies on the previous detection algorithm, which impedes the repair of undetected errors.

In this paper, we propose a novel hybrid data cleaning framework, termed as \textsf{MLNClean}. It aims to address two key challenges: (i) \emph{how to combine the advantages of both quantitative and qualitative techniques to deal with multiple error types}; and (ii) \emph{how to boost cleaning efficiency as much as possible}. %; and (iii) \emph{how to achieve scalability}.
%try to explore some new effective data cleaning strategies in this work.
In the first place, in terms of the first challenge, we seamlessly integrate data quality rules and Markov logic networks (MLNs) into \textsf{MLNClean}, such that it combines the advantages of both qualitative and quantitative techniques, and  it is able to cope with schema-level errors (that violate integrity constraints) and instance-level errors (including replacement errors, typos, and duplicates).
Moreover, we present a critical cleaning criterion based on a new concept of \emph{reliability score}, which is defined by considering both the principle of minimality (using the distance metric) and the statistical characteristics (adopting the weight learning of MLNs).
%Reliability score is composed of two parts, i.e., distance metrics and statistical probability. The distance metrics is used to minimize the cleaning impact on dataset, and the statistical probability is used to make the cleaning results consistent with the statistical distribution. We also integrate deduplicate methods to solve duplicate errors.

Regarding the second challenge, we enable \textsf{MLNClean} to seamlessly handle both error detecting and error repairing stages.
It helps to avoid redundant computation, and therefore minimizes the computation cost of the entire cleaning process.
Furthermore, we develop an effective \emph{MLN index} to shrink the search space.
Specifically, the MLN index is built as a two-layer hash table with each \emph{block} in the first layer including a set of \emph{groups} in the second layer. Each block is with respect to a data rule that involves a set of data attributes. One block contains a set of groups.
In particular, dirty values within each block are cleaned independently, which does not need the access to the information outside the block.
%One group  shares same value(s) on the reasoning part of the corresponding rule w.r.t. the block it belongs to.
%Ideally, if the data are clean, one group contain one piece of attribute values, meaning that same values on reasoning part cannot derive different values on the result part.
%As a result, when one group contains several pieces of attribute values (that are same on the reasoning part), there definitely exist dirty values.
%In this circumstance, we only need to judge which data piece contained in the group is the clean one, and then correct the other dirty ones with the detected clean one, which does not involve the access to the information outside of the group.
In addition, it is noteworthy that, instead of deciding whether each value is clean or not per time in traditional methods, \textsf{MLNClean} chooses to decide whether one piece of data (w.r.t. several attribute values involving one rule) is clean or not per time. Hence, the efficiency of \textsf{MLNClean} is further gained.
In a nutshell, \textsf{MLNClean} has the following contributions.

\begin{itemize}\setlength{\itemsep}{-\itemsep}
%\vspace*{-1mm}
\item Our proposed data cleaning framework \textsf{MLNClean} combines the advantages of both qualitative and quantitative techniques via integrating data quality rules and Markov logic networks (MLNs).
    \textsf{MLNClean} consists of two major cleaning stages, i.e., \emph{cleaning rule-based multiple data versions} and \emph{deriving the unified clean data}, which seamlessly performs error detecting and repairing.
% combines both qualitative and quantitative techniques to clean various types of errors
\item In the first stage, for a data version w.r.t. each block (built in the MLN index), \textsf{MLNClean} first processes \emph{abnormal groups}, and then, it cleans errors within one group using a novel concept of \emph{reliability score}.
 %\textsf{MLNClean} builds an effective MLN index, that is a two-layer hash table with one block in the first layer containing  several groups in the second layer.
    %As a result, \textsf{MLNClean} generates a data version w.r.t. each block (that corresponds to one data rule), using a group merge strategy and a cleaning criterion based on the newly presented concept of \emph{reliability score}.

    %. In particular, the first stage gives the multi-version local optimal cleaning results for the detected errors, the second stage performs the cleaning process on the basis of the previous stage, it unifies the multi-version cleaning results and cleans errors that are not correctly cleaned in the first stage.

\item In the second stage, \textsf{MLNClean} unifies the final clean data set based on multiple data versions in the previous stage, where a newly defined \emph{fusion score} is employed to eliminate conflicts among data versions.

%\textsf{MLNClean} proposes an MLN index construction with \emph{blocking} and \emph{grouping} operations for efficiency enhancement.

%We use MLN to calculate the probability because of the highly fault-tolerant of MLN, which can train a good probability model in case where the dataset contains errors and calculate a reliable probability. Our experiments show that MLNClean can effectively handle the situation where errors amount up to 30\% of total data.

%\item We provide an optimization operation in MLNClean to improve efficiency, i.e., MLN index construction, which contains two operations, Block and Group. Block is to narrow the scope of error detecting. Group is to gather similar dirty data together for the purpose of batch cleaning.

%\item We develop a distributed \textsf{MLNClean} version on the Spark platform to achieve the purpose of \textsf{MLNClean} working well over large-scale datasets.% with a large number of data rules. %And we propose an effective data partition method in order to reduce the time consumption on Markov weight learning.

\item Extensive experiments on both real and synthetic datasets confirm that \textsf{MLNClean} outperforms the state-of-the-art approach in terms of both accuracy and efficiency.
\end{itemize}

The rest of this paper is organized as follows. We review related work in Section~\ref{sec:relatedwork}. Then, Section~\ref{sec:background} introduces data cleaning semantics as well as some concepts related to Markov logic network.
In Section~\ref{sec:framework}, we overview the cleaning framework \textsf{MLNClean}.
Section \ref{sec:cleaning} elaborates the two-stage data cleaning process. Section \ref{sec:spark} details the distributed version of \textsf{MLNClean} on Spark. In Section \ref{sec:experiment}, we report the experimental results and our findings, and then, we conclude our work with future work in Section \ref{sec:conclude}.

\section{Related Work}
\label{sec:relatedwork}

%Data cleaning has become a focus of concern in both academia and industry. The goal of data cleaning is to figure out the most likely erroneous data and give a reliable cleaning method to make it clean afterwards.
%To the best of our knowledge,
Existing data cleaning methods can be partitioned into two categories: (i) qualitative techniques and (ii) quantitative ones.
%Data cleaning methods based on qualitative techniques have hitherto been in the majority.
The qualitative techniques mainly utilize integrity constraints to clean errors,
%. ICs consist of functional dependencies (FDs), conditional function dependencies (CFDs), and denial constraints (DCs). There are plenty of cleaning methods
including ones using FDs \cite{beskales2010sampling, beskales2013relative, bohannon2005cost, kolahi2009approximating}, or CFDs \cite{bohannon2007conditional, cong2007improving, fan2008conditional}, or DCs \cite{bertossi2005complexity, chu2013holistic, lopatenko2007efficient}. %At forepart, data cleaning algorithms can only repair data which violate one specific constraint.
In addition to the above methods that repair data violating only one specific constraint, Temporal \cite{abedjan2015temporal}, LLUNATIC \cite{geerts2013llunatic}, NADEEF \cite{dallachiesa2013nadeef}, BigDansing \cite{khayyat2015bigdansing}, and CleanM \cite{giannakopoulou2017cleanm} support data cleaning against violations of at least two kinds of those constraints.
In particular, Temporal is extended with temporal dimension, to capture the duration information for data cleaning.
%Bart is proposed on top of LLUNATIC that solves data repairing problems.
%LLUNATIC framework solves data repairing problems that involve different kinds of constraints simultaneously.
The generic data cleaning platform NADEEF supports the customization for application-specific data quality problems. It provides a programming interface that allows users to specify multiple types of integrity constraints.
%Subsequently,
% the main contribution of Bart is that it can process large datasets up to million tuples.
BigDansing translates the insights of NADEEF into the Map-Reduce framework. %It employs a series of physical and logical optimizations to enable the distributed cleaning process.
In contrast, the recent work CleanM integrates the physical and logical optimizations used in BigDansing to demonstrate its superiority. It is worthwhile to mention that, CleanM focuses on error detecting regardless of error repairing, but \textsf{MLNClean} considers both  error detecting and repairing.

The quantitative techniques use data itself to construct appropriate models and predict repair solutions on the basis of data distributions. ERACER \cite{mayfield2010eracer}, SCARE \cite{yakout2013don}, and ActiveClean \cite{krishnan2016activeclean} are among this group. ERACER is an iterative statistical framework based on belief propagation and relationship-dependent networks.
%It guarantees accurate inference even dirty data and clean data mixed together.
SCARE cleans data by combining the machine learning and probability models. It repairs errors based on the maximum likelihood estimation.
ActiveClean  is a stepwise cleaning method in which models are updated incrementally rather than retrained, and thus, the cleaning accuracy could gradually increase. The group of methods is more suitable to the cases where the fraction of dirty data is far less than that of clean data. The less the dirty data, the more reliable the learned parameters.
%Conversely, the increase of dirty data fraction makes the learned parameters unreliable, and thus it is difficult to distinguish whether the data are clean.
Theoretically, large-scale datasets can benefit the sophisticated statistical models. %Hence, this category of methods is more adaptable to the increasing amount of data.
%Nevertheless, the cleaning results of them are inferior to qualitative methods.
%In addition, the existing quantitative methods are mostly centralized, and thus have a bottleneck on efficiency. By contrast, \textsf{MLNClean} has a distributed version, which  accelerates the cleaning process without sacrificing cleaning accuracy.

%Both categories of aforementioned cleaning methods have their advantages.

% %some works combine qualitative techniques and quantitative techniques in order to ensure that errors that violates ICs can be cleaned and the cleaning results meet statistical characteristics as well.
One hybrid method \cite{prokoshynaSCMS15} that combines both qualitative and quantitative techniques is then proposed.
However, it only combines FDs and statistical methods without considering other types of integrity constraints.
Thereafter,
%HoloClean \cite{rekatsinas2017holoclean} unifies several data repair signals including different types of ICs to construct a knowledge-base probabilistic graphical models using DeepDive \cite{niu2012deepdive}.
%%HoloClean uses DeepDive to generate errors repair candidates by calculating their marginal probabilities.
%HoloClean is very relevant to our work.
the state-of-the-art method HoloClean \cite{rekatsinas2017holoclean} unifies several data repair signals including integrity constraints and \emph{external} data to construct a knowledge-base probabilistic graphical model by using DeepDive \cite{niu2012deepdive}.
 %, and generates data repair plans based on the probability values of each error value, in other words, it repairs errors once at a time.
HoloClean aims at error repairing, which employs existing approaches to detect errors.
%Hence, two stages of data cleaning are executed in isolation without taking into account the subsequent computation in the cleaning pipeline, incurring redundant computation. Besides, data repairing relies on the previous detection algorithm, which impedes the repair of undetected errors.
By contrast, our proposed framework \textsf{MLNClean} tackle both error detecting and repairing. As empirically confirmed, \textsf{MLNClean} is superior to HoloClean in terms of both efficiency and accuracy.
Also, \textsf{MLNClean} can clean various instance-level errors (e.g., replacement errors and typos), while HoloClean fails to solve them in some cases.
%: (i) introduces an MLN index structure to reduce both search and repair spaces of errors so as to speed up cleaning process, and the experimental results confirm that \textsf{MLNClean} runs faster than HoloClean; (ii) can cope with the situation with various error types of instance level including replacement errors and typos}, while HoloClean fails to solve these erros in some cases (see Section \ref{sec:compared_with_holoClean});
%(iii) has a distributed version on top of Spark platform, with the purpose of accelerating MLN weight learning and achieving scalability, while HoloClean executes data cleaning process in a centralized environment.

\section{Preliminaries}
\label{sec:background}

In this section, we describe data cleaning semantics and some concepts related to Markov logic networks. Table~\ref{tab:symbol} summarizes the symbols used frequently throughout this paper.

\begin{table}
\centering \small
\caption{Symbols and Description}
%\vspace*{-2.5mm}
\label{tab:symbol}
\setlength{\tabcolsep}{3pt}
\begin{tabular}{|c|p{6.7cm}|}
\hline
\textbf{Notation} & \textbf{Description} \\
\hline
$T$ & a dataset with dirty values \\ \hline
$t_i$ & a tuple belonging to the dataset $T$  \\ \hline
$t_i.[A]$ & the value of $t_i$ on attribute $A$    \\ \hline
%$c$ & an attribute value \\ \hline
%$C$ & a set of constants   \\ \hline
%$C(A_j)$ & the value domain of the attribute $A_j$   \\ \hline
$r_i$ & an integrity constraint/rule    \\ \hline
$w_i$ & the weight of a rule $r_i$ \\ \hline
$\gamma$ & a piece of data that contains attribute values of a tuple w.r.t. a rule \\ \hline
$B_i$ & a block (corresponding to a rule $r_i$) in the MLN index over the dataset $T$  \\ \hline
$G_{ij}$ & a group containing a set of $\gamma$s that share the same values on the reason part of the rule w.r.t. the block $B_i$ \\ \hline
%$R$ & a set of integrity constraint rules     \\ \hline
%$L$ & a Markov logic network composed of a set of rule-weight pairs \\ \hline
%$l_i$ & an literal in clausal form   \\ \hline
%$M_{L,C}$ & a Markov network composed of $L$ and $C$\\ \hline
%$gr$ & a ground rule belong to corresponding $r_i$ and $t_i$\\ \hline
\end{tabular}
\vspace*{-2mm}
\end{table}

A dataset \emph{T} with $d$ dimensions $A_1,A_2,...,A_d$ consists of a set of tuples $\{t_1,t_2,...,t_n\}$, and each tuple $t_i$ has the value $t_i.[A_j]\in C(A_i)$ on the attribute $A_i$. $C(A_i)$ denotes the domain of attribute $A_i$.
%can be represent by a series of attribute values. In MLNClean, we denote attributes as $\{A_1,A_2,...,A_m\}$ according to dataset dimension \emph{m}.
%{\color{blue}The smallest data unit that makes up a dataset is constant. Thus, a dataset also can be represented as a series of constants $C = \{c_1,c_2,...,c_n\}$.}
%From the perspective of tuples, a constant $c$ of \emph{i}-th tuple in \emph{j}-th attribute position is denoted as $c=t_i[A_j]$; From the perspective of attributes, the constant $c$ of \emph{j}-th attribute in \emph{i}-th tuple is denoted as $c=A_j[t_i]$, and $A_j[t_i]=t_i[A_j]$.
%Since each attribute has its range of values, take the \emph{j}-th attribute as an example, the range $C(A_j)$ can be denoted as $C(A_j)=\{A_j[t_1],A_j[t_2],...,A_j[t_n]\}$.
There are usually some integrity constraints  that should hold on the dataset $T$, such as functional dependencies (FDs), denial constraints (DCs), as well as conditional functional dependencies (CFDs).
In addition, each integrity constraint could be considered as two parts, i.e., the \emph{reason} part and \emph{result} part, and the \emph{reason} part determines the \emph{result} part.
In other words, there is no the same reason to determine multiple different results.
As an example, for the rule $r_1$, CT is the reason part, while ST is the result part. CT uniquely determines ST.
%As known, each integrity constraint is composed of two parts, i.e., the \emph{reason} part and \emph{result} part,

According to Markov logic theory \cite{domingos2009markov}, every integrity constraint can be converted into a unified form. For ease of presentation, we call the rule in the unified form an \emph{MLN rule}.
The MLN rule is expressed as $l_1 \vee l_2 \vee ...\vee l_n$, where $l_i$ is a literal for $i = 1, \cdots, n$.
A literal is any expression that contains a predicate symbol applied to a variable or a constant, e.g., CT($v$), HN(``ELIZA"). %), PN($t_1.v$) = PN($t_2.v$)}
For the rules shown in Example \ref{example:hai}, they can be transformed into the following MLN rules, where $t_1, t_2\in T$.

%Since integrity constraints is part of first-order logic rules, and first-order logic rules have many different expression forms, we introduce MLN Rules which transform first-order rules into a unified clausal form, thus these rules can be used into Markov logic network.

%Generally, an acceptable implication formula should be in the form: $p_1, p_2, ... , p_m \Rightarrow q_1 \vee q_2 \vee ... \vee q_n$, where $p_i$ and $q_i$ are both literals where the arguments may be either a variable or a constant. Each $p_i$ is a antecedent which must be conjunction, and each $q_i$ is a consequent that must be disjunction.%

%\vspace*{0.05in}
\noindent
\small
\\($r_1$) FD: $\lnot$CT  $\vee$ ST
%\\($r_2$) DC: $\lnot$PN($v_1$) $\vee$ ST($v_2$)
\\($r_2$) DC: $\lnot$(PN($t_1.v_1$) = PN($t_2.v_1$)) $\vee~\lnot$(ST($t_1.v_2$) $\neq$ ST($t_2.v_2$))
\\($r_3$) CFD: $\lnot$HN(``ELIZA")$\vee\lnot$CT(``BOAZ")$\vee $PN(``2567688400")
\vspace*{0.1in}

%\subsection{First-Order Logic}
%
%First-order logic formulas are constructed using the following symbols: constants, variables and predicates. constant symbols refer to $t[A_i]$ in dataset $S$ (e.g., ``ELIZA''). Variable symbols range over the attribute domains, denoted as $v$. And predicate symbols contain attribute names (e.g., CT) in dataset.
%
%First-order logic formulas have different expressions, for instance, if $F_1$ and $F_2$ are formulas, the following are also formulas: $\lnot F_1$ (negation), which is true iff $F_1$ is false; $F_1 \land F_2$ (conjunction), which is true iff both $F_1$ and $F_2$ are true; $F_1 \vee F_2$ (disjunction), which is true iff $F_1$ or $F_2$ is true; $F_1 =>F_2$ (implication), which is true iff $F_1$ is false or $F_2$ is true. Take the ICs mentioned in Example \ref{example:hai} as an example, they are all first-order logic formulas, $r_1$ and $r_3$ are implications, $r_2$ is a negation.

%\subsection{Markov Logic}
%
%Since MLNClean use Markov logic network to unify different expressions of integrity constraints and calculate statistical probabilities. Therefore, we briefly introduce the relevant concepts of Markov logic network here.
%
%Integrity constraints can be expressed as first-order logic rules $R=\{r_1,r_2,...,r_n\}$,

\begin{table}[t]
\centering
\caption{Example of Ground  MLN Rules w.r.t. $r_1$}
\small
\setlength{\tabcolsep}{8pt}
\begin{tabular}{|c|c|c|}
\hline
\multicolumn{1}{|c|}{\textbf{IC rule}} & \multicolumn{1}{c|}{\textbf{MLN rule}} & \multicolumn{1}{c|}{\textbf{Ground MLN rules}} \\ \hline
\multirow{4}{*}{CT $\Rightarrow$ ST} & \multirow{4}{*}{$\lnot$CT $\vee$ ST } & $\lnot$CT(``DOTHAN") $\vee$ ST(``AL") \\
                              &                               & $\lnot$CT(``DOTH") $\vee$ ST(``AL") \\
                              &                               & $\lnot$CT(``BOAZ") $\vee$ ST(``AL") \\
                              &                               & $\lnot$CT(``BOAZ") $\vee$ ST(``AK") \\ \hline
\end{tabular}
\label{table:ground_rules}
%\vspace*{-2mm}
\end{table}

%\begin{figure}[t]
%\centering
%\includegraphics[width=2.2in]{fig/ground_MLN.eps}
%\caption{The ground Markov network w.r.t. $r_1$ over the dataset in Table \ref{table:intro_sample}}
%\label{fig:groun_MLN}
%%\vspace*{-2mm}
%\end{figure}

In a traditional viewpoint of data cleaning work, if a value violates one rule, it has zero probability to be correct.
Nevertheless, in most cases, one cannot guarantee the full correctness of the rules owing to the lack of specific domain knowledge, and hence, it is not desirable to clean data using this kind of \emph{hard} constraints.  %unless the rules are guaranteed to be fully correct.
Fortunately, the attraction of Markov logic networks (MLNs) lies that, it is able to \emph{soften} those constraints.
The formal definition of a Markov logic network is stated below.

%The higher the weight, the greater the difference in log probability between a value that satisfies the rule and one that does not, other things being equal.

\begin{definition}
\label{def:mln}
$($Markov logic network$)$ $\cite{domingos2009markov}$.
A Markov logic network L is defined as a set of rule-weight pairs ($r_i,w_i$), where $r_i$ is a rule and \emph{$w_i$} is a real-number weight of $r_i$.
\end{definition}

%Each MLN rule has an associated \emph{weight} that reflects how strong a constraint is. When a value violates one rule,  it is less probable, but not impossible to be correct. The fewer rules a value violates, the more probable it is to be correct.
Each MLN rule has an associated \emph{weight} that reflects how strong a constraint is. The higher the weight is, the more reliable the rule is, which indicates the higher probability of a value satisfying the rule. Without loss of generality, we use the terms of integrity constraint, rule, and MLN rule interchangeably throughout this paper.

On top of one dataset, each MLN rule can be converted to a set of \emph{ground MLN rules} through a grounding process.
The term \emph{grounding} refers to a process that replaces variables in the MLN rule with the corresponding constants (i.e., attribute values) in the dataset.
%After the grounding process, we can get a series of ground MLN rules and a ground Markov network.
For instance, based on the dataset in Table \ref{table:intro_sample}, the ground MLN rules of the rule $r_1$ are shown in Table \ref{table:ground_rules}. Accordingly, the weight of a ground MLN rule reflects the probability of the attribute values w.r.t. this ground MLN rule being clean.

\section{MLNClean FRAMEWORK}
\label{sec:framework}

\begin{figure}[t]
\centering
%\vspace*{1.5mm}
\includegraphics[width=3.3in]{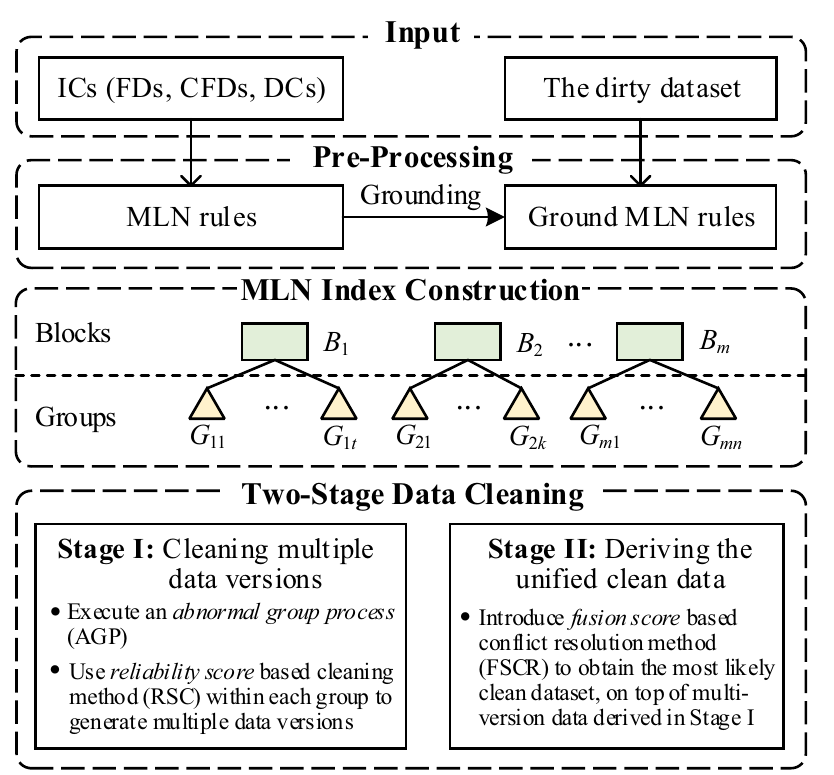}
\vspace*{-2mm}
\caption{\textsf{MLNClean} architecture}
\label{fig:framework}
\vspace*{-4mm}
\end{figure}

In this section, we  briefly introduce the procedure of \textsf{MLNClean}.
 Figure \ref{fig:framework} depicts the general framework of \textsf{MLNClean}.

%\subsection{Overview}
%\label{subsec:overview}

The framework \textsf{MLNClean} receives a dirty dataset together with a set of integrity constraints (ICs). It outputs clean data through \emph{three} steps, including \emph{pre-processing}, \emph{MLN index construction}, and  \emph{two-stage data cleaning}.
%\textbf{Pre-processing}.
In the pre-processing phase, \textsf{MLNClean} first transforms integrity constraints into MLN rules, and derives ground MLN rules of each MLN rule based on the dirty dataset.
%Each MLN rule corresponds to a set of ground MLN rules.
Then, \textsf{MLNClean} builds a two-layer MLN index with a set of \emph{blocks} in the first layer and a set of \emph{groups} in the second layer. The MLN index is a \emph{vital} structure, which helps to narrow the search space of repair candidates for the subsequent data cleaning phase. Next, \textsf{MLNClean} enters the two-stage data cleaning phase, which first cleans \emph{multiple data versions} independently (with each data version coming from different blocks), and then derives the final unified clean data on top of the previous multi-version data. The procedure of \textsf{MLNClean} is shown in Algorithm~\ref{algorithm:framework}.

%Thus, a group of ground MLN rules is obtained.
% which are written in first-order logic, and transform them into MLN rules for a uniform expression. Next, MLNClean performs a grounding operation, which uses constants of the input dirty dataset to replace variables of each MLN rule, we call the grounding result ground rules.

%\subsection{MLN index construction}
%\label{subsec:mln-index}

\textbf{MLN index construction}.
A MLN index is a two-layer hash table. There are a set of \emph{blocks} in the first layer, each of which has a set of \emph{groups} in the second layer. One block corresponds to one MLN rule.
In other words, the data attribute values related to a set of ground MLN rules that belong to one same MLN rule are put together to form one \emph{block}. Thus, the number of blocks is equivalent to the number of MLN rules.
For convenience, we call the data attribute values of each ground MLN rule \emph{a piece of data}, denoted as $\gamma$. % which is the smallest cleaning unit in the following first cleaning stage.
%consists of two main operations: \emph{block} and \emph{group}. We introduce \emph{block} to divide the whole dataset into parts according to rules, each block contains a series of ground rules.
Then, within each block, we get a set of $\gamma$s with the same \emph{reason} part in one \emph{group}.
As a result, in the second layer, each block is divided into several groups, and the pieces of data (i.e., $\gamma$s) within each group share one same reason part (referring to lines 1-13 in Algorithm~\ref{algorithm:framework}).
To be more specific, for different types of rules, we decide the reason and result parts as follows.
First, for implication formulas (i.e., FDs and CFDs), they are in the form: $p_1$, $p_2$, $\dots$, $p_m$ $\Rightarrow$ $q_1$ $\vee$ $q_2$ $\vee\cdots\vee$ $q_n$.
The antecedent belongs to the reason part, while the consequent pertains to the result part.
In contrast, DC formulas have the following form: $\forall~t_1, \cdots, t_k ~\left( p_1 \wedge p_2 \wedge\cdots\wedge p_n \right)$.
We simply treat the last predicate as the result part, and the other predicates as the reason part.

\begin{figure}
\centering
\vspace*{-1mm}
\includegraphics[width=3.4in]{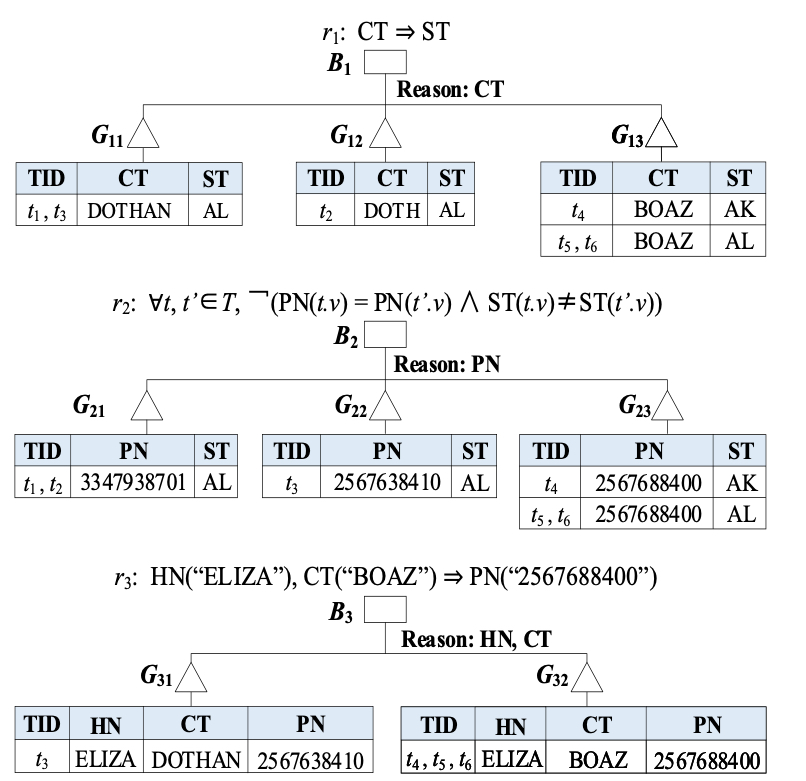}
\vspace*{-2mm}
\caption{Illustration of the MLN index over the sample dataset}
\label{fig:group}
%\vspace*{-10mm}
\end{figure}

Take the sample dataset in Table \ref{table:intro_sample} as an example. We depict the MLN index structure in Figure \ref{fig:group}.
There are three blocks $B_1, B_2$, and $B_3$ corresponding to three rules $r_1, r_2$, and $r_3$ respectively shown in Example~\ref{example:hai}. They have 3, 3, and 2 groups, respectively.
Let $|B|$ and $|T|$ be the number of blocks (or rules) and  tuples in the dataset, respectively. The time complexity of MLN index construction is $O(|B| \times |T|)$. %we traverse each block once to put $\gamma$s with the same reason part into a group, the time complexity of this step is $O(|B| \times |T|)$ as well.
%Thus, the total time complexity of MLN index construction is $O(|B| \times |T|)$.
In addition, it is easy to realize that, there  might be multiple pieces of data pertaining to each tuple in the dataset, and each of them comes from different blocks. In other words, for each tuple, there are at most $|B|$ pieces of data derived from it. Hence, we can say that, there are \emph{multiple data versions}, each of which comes from different blocks.

% the likelihood of candidates being clean data based on the joint probability of GCs by means of MLN weight learning algorithm. MLNClean use Tuffy\cite{niu2011tuffy}, a scaling Markov logic network engine to do MLN weight learning.\\

\begin{algorithm}[t]\small
\label{algorithm:framework}
\LinesNumbered
\DontPrintSemicolon
\caption{The general procedure of \textsf{MLNClean}}
    \KwIn{a dirty dataset $T$ and a set of data rules $R$}
    \KwOut{a clean dataset $T^c$}
        \tcc*{MLN index construction}
        $\mathcal{B} \longleftarrow \varnothing$ \tcp*{$\mathcal{B}$: a block collection}
        \ForEach{$r_i\in R$}
        {
            %distinguish \emph{reason} and \emph{result} parts of $r_i$\;
            $B_i \longleftarrow \varnothing$ \tcp*{$B_i$: a block w.r.t.~$r_i$}
            \ForEach{\emph{tuple $t \in T$}}
            {
                $\gamma_t\longleftarrow \varnothing$ \tcp*{$\gamma_t$: a piece of data for \hspace*{25mm} $t$ w.r.t.~$r_i$}
                $v_l\longleftarrow$ \emph{attribute values} of $t$ w.r.t. the \emph{reason} part of $r_i$\;
                $v_r\longleftarrow$ \emph{attribute values} of $t$ w.r.t. the \emph{result} part of $r_i$\;
                insert $v_l$ and $v_r$ into $\gamma_t$\;
                \If{\emph{there is no group $G_{ij}$ sharing the same $v_l$ from $B_i$}}
                {
               $G_{ij} \longleftarrow \varnothing$ \tcp*{$G_{ij}$: a group}
               add $G_{ij}$ to $B_i$\;
                }
                add $\gamma_t$ to $G_{ij}$ that shares the same $v_l$\;
            }
            insert $B_i$ into $\mathcal{B}$\;
        }
%        \ForEach{$B_i \in \mathcal{B}$}
%        {
%        \ForEach{$\gamma\in B_i$}{
%
%            }
%        }
        % stage1: cleaning within group
        \tcc*{Two-stage cleaning process}
         \ForEach{$B_i \in \mathcal{B}$}
         {
            $B_i \longleftarrow$ \textsf{AGP}($B_i$)\tcp*{abnormal group process}
            \ForEach{\emph{group $G\in B_i$}}{
            \tcp*[l]{R-score based cleaning}
            $G \longleftarrow$ \textsf{RSC}($B_i$, $G$)
            }
            %\ForEach{$G_i \in G$}
            %{
                %$G_i\leftarrow$ Cleaning within Group($G_i$)\;
            %}
         }
        % stage2: conflict resolution
        \tcp*[l]{F-score based conflict resolution}
        $T^c \longleftarrow$ \textsf{FSCR}($T$,~$\mathcal{B}$)
        \Return{$T^c$}
\end{algorithm}

 %errors that are not correctly repaired in the first cleaning stage. In addition, duplicate tuples are removed after eliminating conflicts.

\textbf{Two-stage data cleaning}. %The data cleaning process of \textsf{MLNClean} contains two stages: (i) first cleaning multiple data versions independently, each of which comes from different blocks, and (ii) then deriving the final unified clean data on top of the previous multi-version data, which will be elaborated in Section~\ref{sec:cleaning} later.
For one data version w.r.t. each block (built in the MLN index), the first cleaning stage involves the process of \emph{abnormal groups} (when there are errors located in a rule's reason part), and cleaning the pieces of data (i.e., $\gamma$s) within one group using a new concept of \emph{reliability score} (i.e., r-score).
It refers to  an abnormal
group process strategy (i.e., \textsf{AGP}) and an r-score based cleaning method (i.e., \textsf{RSC}) respectively in lines 14-17 of Algorithm~\ref{algorithm:framework}, which will be elaborated in Section~\ref{sec:cleaning} later.
After cleaning multiple version data independently, there naturally exist conflicts among different data versions.
Thus, in the second cleaning stage, \textsf{MLNClean} strives to eliminate those conflicts using a newly defined concept of \emph{fusion score} (i.e., f-score), in order to get the final clean data.
It is with respect to an f-score based conflict resolution strategy  (i.e., \textsf{FSCR}) in line 18 of Algorithm~\ref{algorithm:framework} (to be detailed in Section~\ref{sec:cleaning}).

%since an error may be included in different rules, we generate multiple versions of local optimal cleaning results for each error based on the rules.
%Specifically, it chooses the clean $\gamma$ based on the reliability score, and replaces other $\gamma$s in the same group with it.
%It is worthwhile to mention that, for a tuple with error(s) in a rule's reason part, the corresponding $\gamma$ might erroneously form a group, and thus resulting in an abnormal group.
%To this end, \textsf{MLNClean} attempts to first identify those abnormal groups and then merge them to corresponding normal groups as correctly as possible.

\section{Two-Stage Data Cleaning}
\label{sec:cleaning}

In this section, we describe the two-stage data cleaning process in \textsf{MLNClean}, including cleaning multiple data versions and deriving the unified clean data.

\subsection{Cleaning Multiple Data Versions}

%Based on the constructed MLN index, each block might have one piece of data w.r.t. a tuple in the dirty dataset.
%Hence, if each piece of data is regarded as one version of the tuple, there are multiple data versions existed in blocks.
As mentioned in Section~\ref{sec:framework}, when cleaning multiple data versions, there are two major missions, i.e.,  \emph{processing abnormal groups} and \emph{cleaning each group}. Through them, schema-level errors (violating integrity constraints) are addressed as well as some instance-level errors including replacement errors and typos.

%In the conventional data cleaning method, the smallest unit is each constant of the input dataset, assuming that there is an error constant $c=t_i[A_j]$, which belongs to the $i-$th tuple in $j-$th attribute position, and a general way to clean it is to find the most likely candidate $c{'}$ from a candidate set $C(A_j)=\{A_j[t_1],A_j[t_2],...,A_j[t_n]\}$. However, the minimal cleaning unit in MLNClean is a ground rule which contains multiple constants.

\subsubsection{Processing Abnormal Groups}

Based on a MLN index, for a tuple with error(s) in the reason part of a data rule, the corresponding piece of data (i.e., $\gamma$) might erroneously form or belong to a group, and thereby, we call the corresponding group \emph{an abnormal group}.
For example, there is a typo $t_2$.[CT] being ``DOTH'' in the sample dataset shown in Table~\ref{table:intro_sample}.
It forms a group $G_{12}$ in the MLN index, as depicted in Figure \ref{fig:group}.
Indeed, the piece of data  being \{CT: DOTH, ST: AL\} w.r.t. $t_2$ should be in the same group $G_{11}$ containing \{CT: DOTHAN, ST: AL\}.
Hence, $G_{12}$ is actually an abnormal group.
To this end, we propose an \emph{abnormal group process} strategy, termed as \textsf{AGP}, to  first identify those abnormal groups and then merge them to the corresponding normal groups.

%For groups in an MLN index, there exist such cases that errors (i.e., typos and replacement errors) appear on the reason part causing those ground MLN rules to be divided into groups incorrectly.
%An abnormality merging strategy is proposed to solve the case that typos appear on the reason part, while replacement errors on the reason part are cleaned by conflict resolution (see Section \ref{sec:conflres}).

%Abnormality merging strategy help abnormal $\gamma$s subsume into correct groups before executing cleaning.
 %for each group $G_{ij}$ within a block $B_i$ in the MLN index, the strategy attempts to determine whether it contains abnormalities.
% for each block $B_i$, it evaluates every group $G_{ij}$ within it.

%{\color{red}We find empirically through experiments that the threshold $\tau$ can be simply set to the $10\%$ of the average of the number of ground MLN rules within group (see Section \ref{experiment:threshold}).

%to represent this whole group, and we treat it as a string, thus the distance between two groups is transformed to the dissimilarity between two $\gamma_1$ and $\gamma_2$.

First, we observe that, the group size and the distance to other groups within the same block are key factors for abnormal group.
In general, the relatively small groups (that have less pieces of data) are prone to be abnormal. The closer to other groups, the most likely to be abnormal. Thus, \textsf{AGP} adopts a simple but effective method to identify abnormal groups. Specifically, if the number of tuples related to $\gamma$s contained in a group is not larger than a threshold $\tau$, \textsf{AGP} regards this group as an abnormal group; otherwise as a normal group. Note that the optimal value of $\tau$ is empirically chosen. Then, for each abnormal group in a block $B_i$, \textsf{AGP} merges it with its nearest normal group within $B_i$.
%Since it is hard to compute the distance between groups,
Specifically, let $\gamma^\star$ be the piece of data related to the most tuples in a group. The distance of two groups is defined as the distance of their respective $\gamma^\star$s.

%In implementation, we employ the Levenshtein distance (LD) and cosine distance (CD).
%The nearest normal group is derived based on the distance.
% functions of Levenshtein distance (LD) and cosine distance (CD).
For instance, in terms of the MLN index shown in Figure \ref{fig:group}, when setting  $\tau$ as 1, groups $G_{12}, G_{22},$ and $G_{31}$ are identified as abnormal groups. Then, for $G_{12}$, its nearest group is $G_{11}$ based on the Levenshtein distance.
Hence, $G_{12}$ is merged with $G_{11}$.
Similarly, $G_{22}$ is merged with $G_{23}$, and $G_{31}$ is merged with $G_{32}$.

The time complexity of \textsf{AGP} is $O(|B| \times |G_a| \times |G-G_a|)$, where $|B|$ is the total number of blocks, and $|G|$ (and $|G_a|$) is the average number of groups (and abnormal groups) in a block.
In addition, it is worth mentioning that, how to identify abnormal groups as accurately as possible is essential to the overall performance of the cleaning framework, since this step has the biggest propagated impact to the final cleaning accuracy. Therefore, we are going to conduct in-depth exploration about this issue in our future work.

\subsubsection{Cleaning within Each Group}

In the constructed MLN index, one group shares the same value(s) on the reason part of the corresponding rule.
 %w.r.t. the block it belongs to.
Ideally, if data are clean, one group contains one and only one piece of data, meaning that the same values on the reason part cannot derive different values on the result part. However, when one group contains several pieces of data (that are the same on the reason part), there definitely exist dirty values. In light of this, we present a cleaning strategy using a  concept of \emph{reliability score},  called r-score based cleaning (\textsf{RSC} for short), to clean dirty values within each group.

\textsf{RSC} judges which piece of data included in the group is clean using the reliability score. Then, \textsf{RSC} corrects the other dirty ones with the detected clean one, so that each group has one and only one piece of data eventually.
It is noteworthy that, \textsf{RSC} cleans the pieces of data within every \emph{block} independently, which does not need the information outside the block. Actually, we can even know that, \textsf{RSC} is executed to clean data within each \emph{group} separately, if regardless of Markov weight learning. In other words, what we would like to highlight here is that, the MLN index structure indeed helps to minimize the search space for \textsf{RSC}.

%Cleaning within group stage aims at cleaning errors which violate ICs rules. Recall that, in an MLN index, $\gamma$s belonging the same group should be identical since no reason can determine multiple different results. Hence, errors exist in the group which contains different $\gamma$s.

The concept of the reliability score is stated in Definition~\ref{def:score}, which is defined to evaluate the possibility degree of the piece of data (i.e., $\gamma$) being clean.
The $\gamma$ with the highest reliability score is most likely clean.
As a result,  all the other pieces of data within the same group are replaced with the piece of data $\gamma$ having the highest reliability score.

\begin{definition}
\label{def:score}
$($Reliability score$)$.
For a piece of data $\gamma_i$ in a group $G$, its reliability score, denoted as r-score$(\gamma_i)$, is defined in Eq. \ref{eq:score}.
\begin{equation}\label{eq:score}
\text{r-score}(\gamma_i)=\min \limits_{\gamma^* \in (G - \{\gamma_i\})} dist(\gamma_i, \gamma^*)\times \Pr(\gamma_i)
\end{equation}
where $dist(\gamma_i, \gamma^*)= \frac{n}{Z}\cdot d(\gamma_i, \gamma^*)$, $Z$ is a normalization function to make $dist(\gamma_i, \gamma^*)$ within the interval $[0, 1.0]$, $n$ denotes the number of tuples related to $\gamma_i$ in group $G$, and $\Pr(\gamma_i)$ is the probability of $\gamma_i$ being clean.
\end{definition}

The definition of reliability score considers two factors: (i) \emph{Distance}, which represents the minimum cost of replacing a $\gamma$ with others. The greater the distance is, the more likely the $\gamma$ is clean. (ii) \emph{Probability}, which indicates the possibility of the $\gamma$ being clean. The higher the probability is, the more likely the $\gamma$ is clean.

%Levenshtein distance quantifies the dissimilarity between $\gamma_1$ and $\gamma_2$ by counting the minimum number of operations required to transform one into the other.
%
%The equation is given by ${lev}_{\gamma_1,\gamma_2}(|\gamma_1|,|\gamma_1|)$ where
%\begin{equation}
%\small
%\begin{aligned}
%{lev}_{\gamma_1,\gamma_2}(i,j)=
%\left\{
%    \begin{array}{lr}
%    \max(i,j)~~~~~~~~~~~~~~~~~~~~~~~~~~{\rm if}\min(i,j)=0\\
%    \min
%    \left\{
%        \begin{array}{lr}
%        {lev}_{\gamma_1,\gamma_2}(i-1,j)+1 &\\
%        {lev}_{\gamma_1,\gamma_2}(i,j-1)+1 &\rm{otherwise.}\\
%        {lev}_{\gamma_1,\gamma_2}(i-1,j-1)+1
%        \end{array}
%    \right.
%    %z=\dfrac{3\pi}{2}(1+2t)\sin(\dfrac{3\pi}{2}(1+2t)), &
%    \end{array}
%\right.
%\end{aligned}
%\label{eq:LD}
%\end{equation}
%
%Cosine distance treats $\gamma_1$ and $\gamma_2$ as vectors, the dissimilarity between them is their cosine of the angle, which is computed as follows
%\begin{equation}
%\begin{aligned}
%d(\gamma_1,\gamma_2)=\frac{\vec {\gamma_1} \cdot \vec {\gamma_2}}{|\vec {\gamma_1}| \cdot |\vec {\gamma_2}|}
%\end{aligned}
%\label{eq:CD}
%\end{equation}

We attempt to derive the probability $\Pr(\gamma_i)$ in Definition~\ref{def:score} as follows.
First, the probability distribution of values $x$ specified by the Markov network is given by Eq.~\ref{equation:markov_prob} from \cite{domingos2009markov}.
\begin{equation}
\label{equation:markov_prob}
\Pr(x)=\frac{1}{Z} \exp \left( \sum_{i=1}^{N}w_i n_i(x) \right)
\end{equation}
where $Z$ is the normalized function, which can be treated as a constant, $n_i(x)$ is the number of true groundings of rule $r_i$ in $x$, $w_i$ is the weight of $r_i$, and $N$ represents the number of rules.
When it computes $\Pr(\gamma_i)$ involving ground MLN rules in our case,
$n_i(\gamma_i)$ equals one for its corresponding ground MLN rule, and $n_j(\gamma_i)$ is zero for other ground MLN rules.
Hence, we have
\begin{equation}
\ln \Pr(\gamma_i)   = w_i - \ln Z
%\ln \Pr(\gamma_i)  = \ln\frac{1}{Z}  \exp(w_i) = w_i - \ln Z
%\begin{aligned}
%\Pr(\gamma)& = \frac{1}{Z} e^{\sum_{i=1}^{N} w_i n_i(\gamma)}   \\
%%   &=\frac{1}{Z} {\rm exp} \left(\sum_{i=1}^{n} w_i f_i(x) \right)   \\
%%   &=\frac{1}{Z} {\rm exp} \left(\sum_{i=1}^{n} w_i \right)~~~~\because f_i(x)=1  \\
%\ln \Pr(\gamma)& = \ln(\frac{1}{Z} \cdot e^{w_i}) = w_i - \ln Z \\
%\end{aligned}
\end{equation}
In particular, $\ln Z$ is a constant, and $\ln\Pr(\gamma_i)$ is a monotonically increasing function. Thus, the higher the probability $\Pr(\gamma_i)$, the greater the weight $w_i$. As a result, instead of deriving the probability $\Pr(\gamma_i)$ directly, we leverage the weight $w_i$ of $\gamma_i$ to compute a reliability score, i.e., r-score$(\gamma_i)=\min_{\gamma^* \in (G - \{\gamma_i\})} dist(\gamma_i, \gamma^*)$ $\times$ $w_i$.

In implementation, the weight $w_i$ is computed by MLN weight learning method from Tuffy \cite{niu2011tuffy}, which adopts diagonal Newton method.
Particularly, the prior weight $w^0_i$ of each $\gamma_i$ for weight learning is given by
\begin{equation}
\begin{aligned}
w^0_i = \frac{c(\gamma_i)}{\sum_{j=1}^{M} c(\gamma_j)}
\end{aligned}
\end{equation}
where $c(\gamma_i)$ represents the number of tuples related to $\gamma_i$, and $M$ denotes the number of different $\gamma$s within a block. For example, for the piece of data being \{CT: BOAZ, ST: AK\} in group $G_{13}$ from block $B_1$, the initial weight is set as $1/6$.

\begin{figure}[t]
\centering
\includegraphics[width=3in]{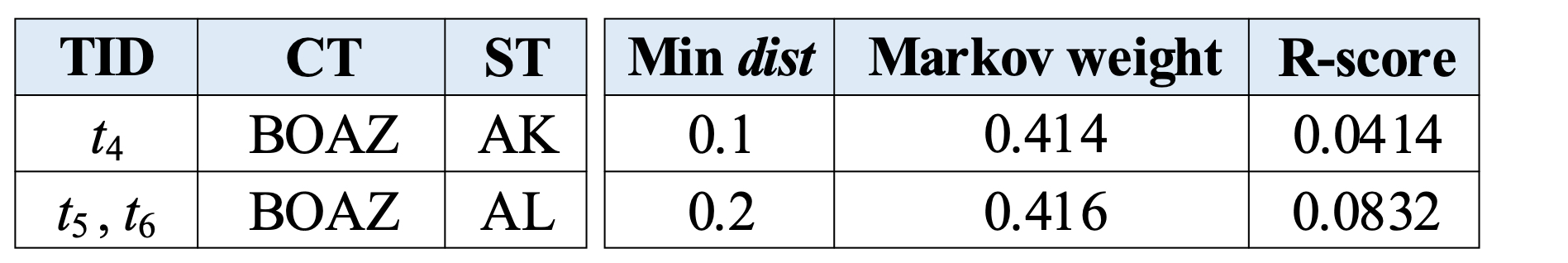}
\vspace*{-1.5mm}
\caption{Illustration of reliability score computation}
\label{fig:group_score}
\vspace*{-4mm}
\end{figure}

% Continuing the previous example, in terms of three blocks in Figure~\ref{fig:group}, there are 3, 3, and 2 groups, respectively.

\begin{example}
Take the group $G_{13}$ belonging to block $B_1$ (as depicted in Figure \ref{fig:group}) as an example. $G_{13}$ includes two pieces of data, denoted as $\gamma_1$ and $\gamma_2$, namely, $\gamma_1$ is \{CT: BOAZ, ST: AK\} and $\gamma_2$ is \{CT: BOAZ, ST: AL\}. They have the same value on the reason part but different values on the result part. Obviously, there is at least one error at the value of attribute  ST  within the group, according to the data rule $r_1$. The reliability score of each $\gamma$ is derived,  as shown in Figure \ref{fig:group_score}, where the Levenshtein distance is used.
%to compute $dist(\gamma_i, \gamma^*)$ in this example, and
The piece of data  $\gamma_1$ being \{CT: BOAZ, ST: AL\} has higher reliability score than $\gamma_2$ being \{CT: BOAZ, ST: AK\}. Thus, $\gamma_1$ is regarded as the clean one in this group, and  $\gamma_2$ is finally replaced with $\gamma_1$ by \textsf{RSC} strategy.

\begin{figure*}[t]
\centering
%\hspace*{1.5mm}
\includegraphics[width = 7.5in]{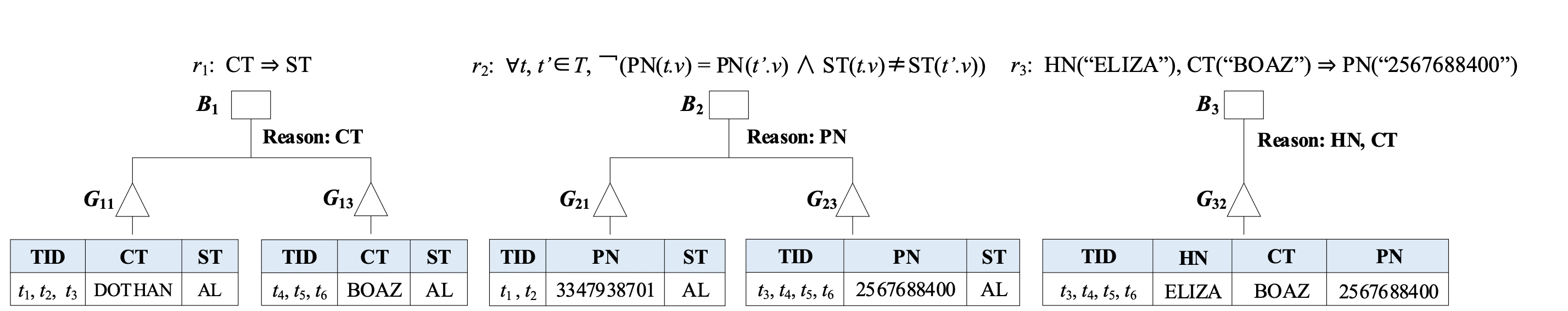}
\vspace*{-4mm}
\caption{The three clean data versions over the sample dataset}
\label{fig:abnormal_merging}
\end{figure*}

In addition, for ease of understanding, Figure \ref{fig:abnormal_merging} illustrates three clean data versions after adopting \textsf{AGP} and \textsf{RSC} strategies consecutively.
%content of groups with reliability scores after abnormality merging in Figure \ref{fig:abnormal_merging}.
In particular, the group $G_{21}$ from $B_2$ is skipped to calculate the reliability score, because this group has reached the ideal state with only one $\gamma$ in it. Finally, the first clean data version contains \{CT: DOTHAN, ST: AL\} in $G_{11}$ (w.r.t. $t_1, t_2,$ and $t_3$) and \{CT: BOAZ, ST: AL\} in $G_{13}$ (w.r.t. $t_4, t_5,$ and $t_6$). The second clean data version incorporates \{PN: 3347938701, ST: AL\} in $G_{21}$  (w.r.t. $t_1$ and $t_2$) and \{PN: 2567688400, ST: AL\} in $G_{23}$ (w.r.t. $t_3, t_4, t_5,$ and $t_6$). The third clean data version consists of \{HN: ELIZA, CT: BOAZ, PN: 2567688400\} in $G_{32}$ (w.r.t. $t_3, t_4, t_5,$ and $t_6$).
\end{example}

\subsection{Deriving the Unified Clean Data}
\label{sec:conflres}

The first cleaning stage of \textsf{MLNClean} has obtained the ruled-based multiple clean data versions.
Now, we are ready to enter the second cleaning stage. It aims to execute the data fusion on top of multi-version data, in order to get the final clean data, in which data conflicts among different data versions have to be solved.
Note that, this stage provides the second opportunity  to clean erroneous data that are not  (or incorrectly) repaired   in the first cleaning stage as many as possible.

%Conflict resolution stage aims to achieve two goals: (i) to unify the results of the multiple versions of the cleaning produced in the first phase; (2) to clean replacement errors which are not properly cleaned in the first cleaning stage.

Take the tuple $t_3$ depicted in Table \ref{table:intro_sample} as an example. After finishing the first cleaning stage, the piece of data w.r.t. $t_3$ in $B_1$ (w.r.t. the first data version) is \{CT: DOTHAN, ST: AL\}, while in $B_3$ (w.r.t. the third data version), the piece of data related to $t_3$ is \{HN: ELIZA, CT: BOAZ, PN: 2567688400\}.
It is obvious that, $t_3$.[CT] has two different values (i.e., ``DOTHAN'' and ``BOAZ'') from the two versions.
That is to say, there exist conflicts on attributes  CT of  $t_3$, and  that should be eliminated to get the final clean $t_3$.
Besides, although \{CT: DOTHAN, ST: AL\} conforms to the rule $r_1$, there might still exist errors w.r.t $t_3$ in the case that it is erroneously classified into a group and thereby cannot be correctly repaired in the first cleaning stage.

Accordingly, during executing the data fusion on top of multiple data versions, we identify a set of \emph{candidates} to solve conflicts, which refers to all the possible fusion versions for a tuple. Hence, we introduce a novel concept of \emph{fusion score} (i.e., f-score) to get the most likely clean fusion version.
%For simplicity, we call it f-score based conflict resolution (\textsf{FSCR} for short) strategy.
Specifically, for a tuple $t$, the \emph{fusion score} of $t$, denoted by f-score($t$), is defined as the product of weights of data pieces $\gamma^1, \cdots, \gamma^m$ (related to $t$) from different data versions, as written in Eq. \ref{equation:combine}.
\begin{equation}
\label{equation:combine}
\text{f-score}(t)  = w_{1}\times  \cdots \times w_{m}
\end{equation}
where $w_i$ denotes the weight of $\gamma_i$ related to $t$. The larger the f-score, the more likely clean the corresponding fusion version of tuple $t$.
{\color{red}   }

\begin{algorithm}[t]\small
\label{algorithm:conflict_resolution}
\LinesNumbered
\DontPrintSemicolon
\caption{F-Score based Conflict Resolution (\textsf{FSCR})}
    \KwIn{a dirty dataset $T$ and multiple clean data versions contained in blocks}
    \KwOut{a clean dataset $T^c$}
    $T^c\longleftarrow\varnothing$\;
        \ForEach {\emph{tuple $t\in T$}}
        {
            $\mathcal{V}(t) \longleftarrow \{\gamma_t^1, \cdots, \gamma_t^{m}\}$\; %identify conflicts for $t_i$ in each block;\\
            %\If(\tcp*[h]{No conflict exists}){$\upalpha=\varnothing$}
            %{
            %    break;\\
            %}
            $f_{max}\longleftarrow 0$; $t_{f_{max}}\longleftarrow t$\; %$cter\longleftarrow 0$\;
            \ForEach{$\gamma_t^i$ \emph{from} $\mathcal{V}(t)$}
{
$f\longleftarrow w_{\gamma_t^i}$; $\mathcal{V}(t)\longleftarrow \mathcal{V}(t)-\{\gamma_t^i\}$\;
$\langle t_{f_{max}}, f_{max}\rangle\longleftarrow$\textsf{GetFusionT}($\mathcal{V}(t), f, f_{max},  \gamma_t^i$)\;
%\If{$f > f_{max}$}
%                {
%                    $t_{f_{max}} \longleftarrow \gamma_t^i$; $f_{max} \longleftarrow f$\;
%                }
}
             %$t_{f_{max}}\longleftarrow$\textsf{GetCleanT}($\mathcal{V}(t), f_{max}$)\;
             %$t\longleftarrow t_{c_{max}}$\;
             replace the corresponding attribute values of $t$ with $t_{f_{max}}$\;
             add $t$ to $T^c$\;
        }
\Return{$T^c$}

\textbf{Function} \textsf{GetFusionT}($\mathcal{V}, f, f_{max},  \gamma_t^i$)\;
%select a piece of data $\gamma_t^j$ from $\mathcal{V}(t)$\;
%\If{$f = 0$}
%{\Return{$\langle t_{f_{max}}, f_{max}\rangle$}\;}
\If{\emph{$\mathcal{V}$ is empty} $\mid\mid f = 0$}
{
\If{$f > f_{max}$}
                {
                    $t_{f_{max}} \longleftarrow \gamma_t^i$; $f_{max} \longleftarrow f$\;
                }
 \Return{$\langle t_{f_{max}}, f_{max}\rangle$}
}

 \ForEach{$\gamma_t^j$ \emph{from} $\mathcal{V}$}
 {
$\mathcal{V} \longleftarrow \mathcal{V} - \{\gamma_t^j\}$\;
 %$\gamma \leftarrow$ the attribute values of $t_i$ in $B_j$;\\
                    \If{\emph{$\gamma_t^i$ and $\gamma_t^j$ have conflicts}}
                    {
                    \If{\emph{there exists $\gamma{'} \in B_j - \{\gamma_t^j\}$ with the highest $w_{\gamma{'}}$ such that there is no conflict between $\gamma_t^i$ and $\gamma{'}$}}
                    {
                          $\gamma_t^j\longleftarrow \gamma{'}$\;
                          }
                        \Else
                        {
                            $f\longleftarrow 0$; \Return{$\langle t_{f_{max}}, f_{max}\rangle$}\;
                        }

                    }
                    $\gamma_t^i \longleftarrow \gamma_t^i \cup \gamma_t^j$;
                        $f\longleftarrow f \times w_{\gamma_t^j}$\;
%\textsf{GetCleanT}($\mathcal{V}(t), f_{max}, cter, \gamma_t^i$)\;
\textsf{GetFusionT}($\mathcal{V}, f, f_{max},  \gamma_t^i$)\;}
\Return{$\langle t_{f_{max}}, f_{max}\rangle$}
\end{algorithm}

%We develop a conflict resolution strategy to execute data fusion on top of multiple data versions.
%The strategy leverages a newly introduced concept of

As a result, we develop an f-score based conflict resolution (\textsf{FSCR} for short) strategy, with the pseudo code presented in Algorithm \ref{algorithm:conflict_resolution}. % gives  of \textsf{FSCR}.
It receives a dirty dataset $T$ and multiple clean data versions obtained in the first cleaning stage, which are stored in different blocks. To begin with, \textsf{FSCR} initializes the clean dataset $T^c$ as an empty set (line 1).
Then, for each tuple in the dirty dataset $T$, it attempts to derive the unified clean tuple (lines 3-9).
\textsf{FSCR} first puts all the pieces of data  $\gamma_t^1, \cdots, \gamma_t^m$ into a set $\mathcal{V}(t)$, where $m$ is not larger than $|B|$ (line 3), i.e., $\mathcal{V}(t)$ collects all the versions of tuple $t$.
Next, a temporal variable $f_{max}$ is set as zero, which is used to store the maximal f-score, and the corresponding fusion version of $t$, i.e., $t_{f_{max}}$, is set as $t$ (line 4). Then, for each piece of data $\gamma_t^i$ related to $t$,
\textsf{FSCR} merges it with other data pieces of $t$ contained in $\mathcal{V}(t)$ using the function \textsf{GetFusionT}, in order to find the optimal fusion version with the highest f-score as the final unified one (lines 5-7).
Thereafter, the tuple $t$ is updated using the derived optimal fusion version $t_{f_{max}}$, and is added to $T^c$
(lines 8-9). \textsf{FSCR} proceeds to process the remaining tuples in the dirty dataset $T$ one by one. Finally, it returns the clean dataset $T^c$ (line 10).

Since the fusion version of a tuple is related to the order of merging $\gamma_t^i$ ($i = 1, \cdots, m$), the number of possible fusion versions is a factorial number, up to $m!$.
Hence, \textsf{GetFusionT} is a recursive function (lines 11-25).
It terminates if f-score is zero, or one fusion version of tuple $t$ has been obtained (i.e., $\mathcal{V}$ becomes empty) (lines 12-15).
Given a $\gamma_t^i$, for each  data piece $\gamma_t^j$ from  $\mathcal{V}$ (that excludes $\gamma_t^i$), \textsf{GetFusionT} has to decide whether there are conflicts between $\gamma_t^i$ and $\gamma_t^j$.
The conflicts exist only in the case that  $\gamma_t^i$ and $\gamma_t^j$ have some common attribute(s), but the values on at least one of those attribute(s) from  $\gamma_t^i$ and $\gamma_t^j$ are different.
%To be more specific, for two pieces of data $\gamma_t^i$ and $\gamma_t^j$, \textsf{FSCR} complies with the following three principles.
%(i) If $\gamma_t^i$ and $\gamma_t^j$ do not share any attribute, there is definitely no conflict;
%(ii) If $\gamma_t^i$ and $\gamma_t^j$ have some common attribute(s), and the values on each of those attribute(s) from  $\gamma_t^i$ and $\gamma_t^j$ are identical, there is no conflict;
%and (iii)
If existing conflicts, \textsf{GetFusionT} attempts to find a candidate piece of data $\gamma{'}$ from the block $B_j$  to replace $\gamma_t^j$. The candidate $\gamma{'}$ is the one with the highest Markov weight, and has no conflict with $\gamma_t^i$.
%Namely, $\gamma{'}$ $\gamma_t^i$.
If there does not exist such $\gamma{'}$, the fusion for tuple  $t$ fails and terminates (lines 18-22).
If there is no conflict or exists a proper $\gamma{'}$, \textsf{GetFusionT} continues the tuple fusion, and meanwhile updates the value of f-score (line 23). Note that, in line 6, the value of f-score is firstly set as the weight of current $\gamma_t^i$.
In the sequel, it invokes itself at line 24 to merge with the other version of $t$.

\begin{example}
We illustrate how \textsf{FSCR} works in terms of tuple $t_3$ in the sample dataset, based on three versions (denoted by $\gamma_1, \gamma_2,$ and $\gamma_3$) of tuple $t_3$ (obtained from the first cleaning stage).
In particular, $\gamma_1$ denotes  \{CT: DOTHAN, ST: AL\} from block $B_1$, $\gamma_2$ is with respect to \{PN: 2567688400, ST: AL\} from $B_2$, and  $\gamma_3$ represents \{HN: ELIZA, CT: BOAZ, PN: 2567688400\} from block $B_3$.
Hence, following Algorithm~\ref{algorithm:conflict_resolution}, for the tuple $t_3$, \textsf{FSCR} gets $\mathcal{V}(t_3) = \{\gamma_1, \gamma_2, \gamma_3\}$. For simplicity, we show two fusion attempts: (i) merging $\gamma_1$, $\gamma_2$, and  $\gamma_3$ in order, and (ii) merging $\gamma_2$, $\gamma_3$,  and $\gamma_1$ in order.
%\textsf{FSCR} starts to the fusion attempts for tuple $t_3$.
For the first attempt, $\gamma_1$ and $\gamma_2$ are merged directly, since there is no conflict between them.
Then, it proceeds to merge with $\gamma_3$ from block $B_3$. While there is a conflict on the attribute CT, %In the first case, we use $\alpha_1 \in B_1$ as the benchmark to combine with the $\gamma$ w.r.t $t_3$ in $B_2$, since there is no conflict we combine them directly and the benchmark is updated to $\alpha_i$=\{CT: DOTHAN, ST: AL, PN: 2567688400\}, then we continue to combine another $\gamma$ related to $t_3$ in $B_3$, we find this $\gamma$ and the updated benchmark have conflict in `CT' attribute.
here \textsf{GetFusionT} tries to find another $\gamma'$ from block $B_3$, such that $\gamma'$ has the same value on the common attribute CT for the fusion of $\gamma_1$ and $\gamma_2$.
%Namely, \textsf{FSCR} has to decide whether there is a $\gamma'$ in block $B_3$ having the value ``DOTHAN'' on the attribute CT.
%Nevertheless,
Unfortunately, there does not exist such $\gamma'$ in block $B_3$. Thus, \textsf{GetFusionT} terminates the current fusion.
For the second attempt, $\gamma_2$ and $\gamma_3$ are firstly merged, as there is no conflict between them.
Then, \textsf{GetFusionT} merges the fusion of $\gamma_2$ and $\gamma_3$  with $\gamma_1$ from block $B_1$.
However, there is a conflict on the attribute CT. At that time, \textsf{GetFusionT} successfully finds a $\gamma'$ being  \{CT: BOAZ, ST: AL\} from block $B_1$. Finally, the final fusion version of $t_3$, i.e., \{HN: ELIZA, CT: BOAZ, ST: AL, PN: 2567688400\}, is obtained.
\end{example}

%for each $ct_i\in CT$, at first we use the ground rule which contains $ct_i$ as a benchmark to match another GC with the same value as $ct_i$ in other domain, and use the combined GC as a new benchmark to combine with other GCs in remaining domains. Before each combine operation, we need to check if the benchmark has the same attribute with the remaining domain to be merged. If yes, find the one which has the same value with the benchmark in this domain, and combine it to re-generate a larger combined benchmark. Otherwise, combine it directly. After all domain is merged, we get a complete tuple as a possible combination result. Ultimately, all candidate possible combination result is found after go through all $ct_i\in CT$.

%The overall algorithm is shown in Algorithm \ref{algorithm:combine}.

Let $|T|$ be the number of tuples in the dataset and $m$ be the average number of versions for a tuple. $m$ is bounded by the number of blocks/rules. There are at most $m!$ fusion versions for a tuple. Each fusion version needs $O(m)$ time to find the candidate version $\gamma'$. Therefore, \textsf{FSCR} takes $O(|T| \times m! \times m)$ time. %, where $|G|$ is the average number of groups  in a block.
In addition, after eliminating conflicts via \textsf{FSCR}, \textsf{MLNClean} automatically detects and removes duplicate tuples. Take the sample dataset as an example, $t_1$ and $t_2$ are duplicates, and $t_3, t_4, t_5,$ and $t_6$ are duplicates. \textsf{MLNClean} deletes extra duplicate tuples among them to finish the cleaning process.

\section{Distributed MLNClean}
\label{sec:spark}

%Most of the existing data cleaning techniques containing machine learning methods \cite{krishnan2016activeclean, mayfield2010eracer, rekatsinas2017holoclean, yakout2013don} are centralized. We show that MLNClean can be scalable by implementing it as a Spark application.

%If the data set is large or the size of data rules increases
In order to enable \textsf{MLNClean} to work well even for large-scale datasets with a large number of data rules, we aim to deploy \textsf{MLNClean} in the Spark system. In this section, we describe the distributed \textsf{MLNClean} program.

%As shown in Algorithm \ref{algorithm:spark_overall}, at first a data partition strategy is proposed to divide the whole dataset into small partitions (Line 1), and then we construct MLN indexes for each partition, where Markov weight learning is implemented to get the most likely weight of each $\gamma$. Whereafter, the weight of each $\gamma$ is adjust to ensure the reliability (Line 2). Then, we execute the two-stage cleaning process in each partition (Line 3) and remove duplicate tuples in the end (Line 4). And we illustrate each step in details as follows.

%In scenarios of
%As analyzed, Markov weight learning dominates the computation cost in \textsf{MLNClean}.
 %a significant factor that limits scalability is the complexity of MLN weight learning.
%Since each MLN rule consists of multiple constants, if the value range of data enlarges, the number of MLN rules will increase in unpredictable circumstances.
%The weight learning process will become much slower in result.
%Tuffy finds empirically that a well sampled smaller network can achieve similar weights as a larger world using much less time \cite{doan2011user}.

%Since sufficient evidences can achieve optimal weights of MLN so as to get well sampled networks,

First, data skew is a critical issue in the distributed system, which may lead the overall process to delay. Hence, an effective data partition strategy is needed for the distributed \textsf{MLNClean} version.
As a result, the distributed \textsf{MLNClean} version executes in the following procedure.
It first partitions the whole dataset into several parts, and allocates each part to a worker node. Then, it cleans each part using the stand-alone \textsf{MLNClean}. When each part has been cleaned, those parts are gathered to derive the final clean dataset, during which conflicts and duplicates are eliminated in the same way to stand-alone  \textsf{MLNClean}.

%As known, data skew may lead to some RDD operation executing for a long time, making the overall process delay.
%Hence, we introduce an effective data partition algorithm, which is responsible for partitioning the dataset into $k$ partitions {\color{red}as uniformly as possible and ensuring} similar tuples belong to the same partition.
%Specifically, our data partition strategy assigns tuples to each partition as uniformly as possible, in order to avoid or alleviate the data skew issue.
%Meanwhile, we also try to make the number of tuples uniformly distributed in each cluster as much as possible in order to achieve the uniform distribution target, thereby reducing the disadvantages caused by data skew, i.e.,

\begin{algorithm}[t]\small
\label{algorithm:spark_partition}
%\SetNlSty{small}{}{:}
%\SetNlSkip{4mm}
\LinesNumbered
\DontPrintSemicolon
\caption{Data partition method}
    \KwIn{a dataset $T$, a data partition $\mathcal{P} = (P_1, \cdots, P_k)$}
    \KwOut{the data partition $\mathcal{P}$}
        $s\longleftarrow\lceil \frac{|T|}{k} \rceil$\;
        initialize each part $P_i$ as an empty max-heap for $i = 1, \cdots, k$\;
        randomly select a centroid $o_i$, and insert it into the part $P_i$ for $i = 1, \cdots, k$\;
        collect those $k$  centroids into $O$\;
        %insert each centroid from $O$ into the corresponding partition\;
        %initialize centroids of clusters $O$ as $k$ tuples from $T$\;
        %initialize each max heap $P_j$ with the maximum size $\lceil \frac{|T|}{k} \rceil$;\\
        %$O \leftarrow$pickCenter($T$, $k$); // Randomly select center tuples\\
%        \For{\emph{each $o_{j}\in O$}}
%        {
%            $P_j$.insert($o_j$)\;
%        }
        \For{\emph{each $t_i\in (T - O)$}}
        {
            find $P_j$ with $\min \limits_{o_j \in O} dist(t_{i}, o_j)$\;
%            minDis = $\infty$;\\
%            $\lambda$ = 0;\\
%            \For{$each\;o_{j}\in O_j$}
%            {
%                Dis = distance($t_i$, $o_j$)\;
%                \If{Dis $\leqslant$ minDis}{
%                    minDis = Dis;\\
%                    $\lambda$ = $j$;\\
%                }
%            }
            \If{$|P_j| < s$}
            {
                insert $t_i$ into the part $P_j$\;
            }
            \Else
            {
                $t_{e} \longleftarrow t_i$; get the top node $t_{top}$ of $P_j$\;
                \If{$dist(t_i, o_j)$ $<dist(t_{top}, o_j$)}
                {
                    $t_{e} \longleftarrow t_{top}$; replace $t_{top}$ with $t_i$\;

                }
                find $P_t$ with $\min \limits_{o_t \in \mathcal{O}, |P_t|< s} dist(t_{e}, o_t)$\;
                add $t_{e}$ to $P_t$\;
            }

        }
        \Return{$\mathcal{P}$}
\end{algorithm}

\begin{figure}[!]
\centering
%\vspace*{-2mm}
\includegraphics[width=2in]{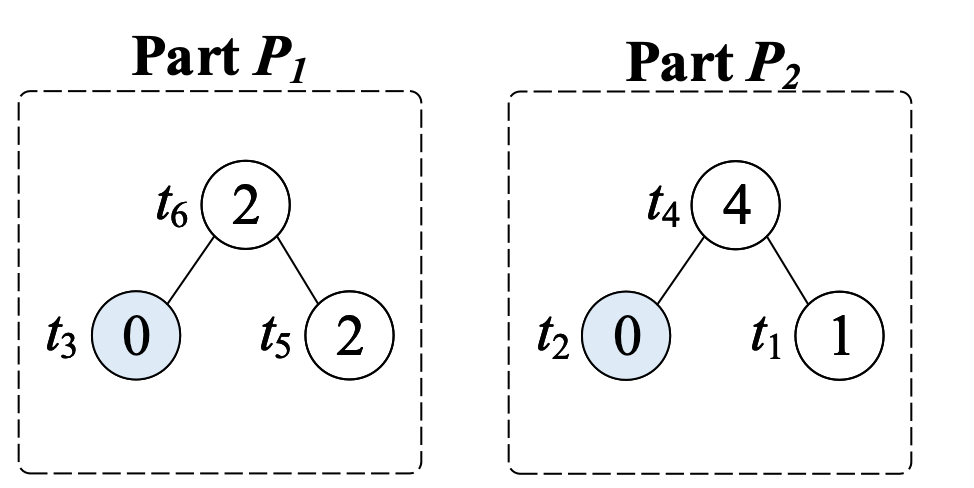}
\vspace*{-1.5mm}
\caption{Illustration of data partition}
\label{fig:spark-partition}
\vspace*{-4mm}
\end{figure}

Our data partition method is depicted in Algorithm \ref{algorithm:spark_partition}, which aims to divide
 the dataset $T$  into $k$ parts, denoted as $\mathcal{P} = \{P_1,$ $P_2,$ $\cdots,$ $P_k\}$.
  %with centroids $O = \{o_1, o_2, \cdots, o_k\}$.
%Every centroid is with respect to a tuple $k$ selected randomly.
% as centroids $O = \{o_1, o_2, \cdots, o_k\}$.
The maximum capacity of each part is set as   $\lceil \frac{|T|}{k} \rceil$ (denoted by $s$) for uniform distribution (line 1).
Tuples are stored in a maximum heap for each part, where each node also records the distance between the corresponding tuple and the centroid of the part.
%Then, we randomly select $k$ tuples from the dataset as the centroid for each partition, and put them into their corresponding partitions (line 2-4).
Then, each tuple $t_i$ is allocated to the part $P_j$, if the distance between $t_i$ and $o_j$ is minimal and the number of tuples in $P_j$ is smaller than $s$ (lines 6-8).
%Specifically, there exists such a case that the cluster is full when putting a tuple into it, i.e.,
When the size of $P_j$ is larger than $s$, if the distance $dist(t_i, o_j)$ is smaller than the distance from the top node (denoted by $t_{top}$) to the centroid of this part $P_j$ (i.e., $dist(t_{top}, o_j)$),
$t_{top}$ is replaced with $t_i$ in the part $P_j$.
Meanwhile, $t_{top}$ is inserted into its closest partition $P_t$ if it is not full.
% is farther than the distance recorded by the current tuple $t_i$, we put $t_i$ into the master node, and put the old $t_{top}$ into another closest cluster $P{'}$ which is not full;
On the other hand, if $dist(t_i, o_j)\geq dist(t_{top}, o_j)$, $t_i$ is directly added to its closest part $P_t$ that is not full (lines 10-14).
%We repeat above steps until all tuples are classified into clusters.

Take the sample dataset shown in Table \ref{table:intro_sample} as an example. Figure \ref{fig:spark-partition} illustrates the partition result. The inner digital of each part denotes the distance between the tuple and the centroid. For instance, the digital 4 of partition $P_2$ represents the distance between tuple $t_4$ and the centroid $t_2$.

The time complexity of data partition algorithm is $O(|T| \times \lg s)$, where $|T|$ is the total number of tuples in the dataset, and $O(\lg s)$ is the complexity of one insertion operation in a maximum heap.

%\begin{example}
%We review the dataset of Table \ref{table:intro_sample}. Given the partition number $k=2$, the partition result according to $\mathcal{P}$ as shown in Figure \ref{fig:spark-partition}(a), we can see that $p_1=\{t_1,t_2\}$ and $p_2=\{t_3,t_4,t_5,t_6\}$, and it cause data skew because the tuple number is not equal between the two partitions. Thus we should execute repartition operation in order to achieve load balancing. We calculate the average tuple number in each partition as $\overline{n}$, and then allocate $\overline{n}$ tuples for each partition, we randomly assign the extra tuples to the insufficient number of partitions. In this example, $\overline{n}=3$, and we reassign $t_6$ from $p_2$ to $p_1$ as shown in Figure \ref{fig:spark-partition} (b).
%\end{example}

%As a result, when having gotten the data partition using Algorithm~\ref{algorithm:spark_partition}, the cleaning of tuples in each {\color{red}partition} can be performed by the stand-alone \textsf{MLNClean} version.\

For the distributed \textsf{MLNClean} program, due to the small-scale of tuples allocated in each part, the result of every weight learning on each worker (w.r.t. each part) might not be very reliable. %it needs to construct a relatively small Markov network on each part so as to execute the Markov weight learning method in parallel.
For example, in the part $P_2$, the learned weight of $\gamma$ being \{CT: DOTHAN, ST: AL\} may be unreliable, since there is no relevant evidence for learning the weight of $\gamma$. Conversely, in the part $P_1$, there is a tuple $t_3$ providing the relevant evidence for learning the weight of $\gamma$. Thus, we adjust the weight of each $\gamma$  via Eq.~\ref{eq:weight} in virtue of the relevant evidence from other parts.
\begin{equation}~\label{eq:weight}
\overline{w}(\gamma) = \frac{\sum_{i = 1}^{k}n_{_i} \times w_i}{\sum_{i = 1}^{k}n_i}
\end{equation}
where $k$ is the number of parts in the data partition, $n_i$ is the number of tuples related to $\gamma$ in the part $P_i$, and $w_i$ represents the learned weight of $\gamma$ in the part  $P_i$.
As a result, each $\gamma$ corresponds to a unique weight $\overline{w}$ in global, and  it is used for the subsequent cleaning process.

%In addition, it is necessary to mention that, there exist duplicates after finishing the cleaning process in each partition.
%Thus, as the final step, \textsf{MLNClean} scans the dataset to find duplicate tuples and remove them, in order to complete the whole data cleaning process on Spark platform.

%the last operation is deduplication, for the reason that

%we divide the whole dataset into several partitions, there may still exists duplicates after each partition finished its own deduplicate operation.
%To solve this problem,

%\begin{table}[t]
%\centering
%\caption{Datasets information}
%\label{table:dataset_info}
%\begin{tabular}{|l|c|c|c|}
%\hline
%\textbf{Parameter} & \multicolumn{1}{l|}{\textbf{HAI}} & \multicolumn{1}{l|}{\textbf{CAR}} & \multicolumn{1}{l|}{\textbf{TPC-H}} \\ \hline
%Rows               & 231,265                           & 30,760                            & 6,001,115                           \\
%rules number       & 7                                 & 2                                 & 1                                   \\ \hline
%\end{tabular}
%\end{table}

\section{Experiments}
\label{sec:experiment}

In this section, we present a comprehensive experimental evaluation.
In what follows, we evaluate our proposed data cleaning framework \textsf{MLNClean} using both real-world and synthetic datasets in the following scenarios: (i) the experimental comparisons between \textsf{MLNClean} and the state-of-the-art method HoloClean \cite{rekatsinas2017holoclean}, (ii) the effect of various parameters on the performance of \textsf{MLNClean},  and (iii) the performance of distributed \textsf{MLNClean} version on the Spark platform.

\newcommand{\tabincell}[2]{\begin{tabular}{@{}#1@{}}#2\end{tabular}}  %±í¸ñ×Ô¶¯»»ÐÐ
\begin{table}[t]
\centering\small
\vspace*{-2mm}
\caption{Rules Used in Each Dataset}
\vspace*{0.5mm}
\setlength{\tabcolsep}{10pt}
\label{table:Dataset Rules}
\begin{tabular}{|l|l|}
\hline
\textbf{Dataset}     & \textbf{Rules}                                                                                \\ \hline
\multirow{7}{*}{\emph{HAI}} & PhoneNumber $\Rightarrow$ ZIPCode                                                            \\
                     & PhoneNumber $\Rightarrow$ State                                                              \\
                     & ZIPCode $\Rightarrow$ City                                                                   \\
                     & MeasureID $\Rightarrow$ MeasureName                                                                  \\
                     & ZIPCode $\Rightarrow$ CountyName                                                             \\
                     & ProviderID $\Rightarrow$ City, PhoneNumber                                                    \\
                     & \tabincell{l}{$\forall t, t',\lnot$(PhoneNumber($t.v$) = PhoneNumber($t'.v$)\\ $\land$ State($t.v$)$\neq$ State($t'.v$))} \\ \hline
\multirow{2}{*}{\emph{CAR}} & Make{(}``acura"{)}, Type $\Rightarrow$ Doors                                                   \\
                     & Model, Type $\Rightarrow$ Make                                                                \\ \hline
\emph{TPC-H}                & CustKey $\Rightarrow$ Address                                                                \\ \hline
\end{tabular}
%\vspace*{-2mm}
\end{table}

\subsection{Experimental Setup}

\begin{figure*}[t]
\centering
\includegraphics[width=0.32\textwidth]{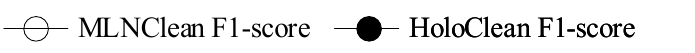}\includegraphics[width=0.3\textwidth]{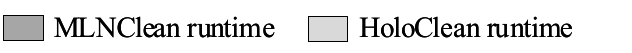}\\
\vspace*{-3mm}
\hspace*{-6mm}
\subfigure[\emph{CAR}]{
   \includegraphics[height=1.1in]{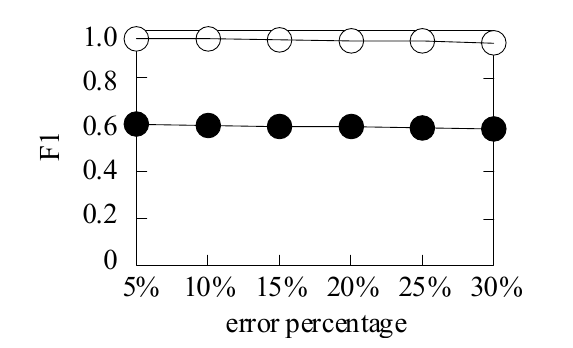}\label{fig:e-a-car}
  }%\hspace*{-8mm}
\subfigure[\emph{HAI}]{
   \includegraphics[height=1.1in]{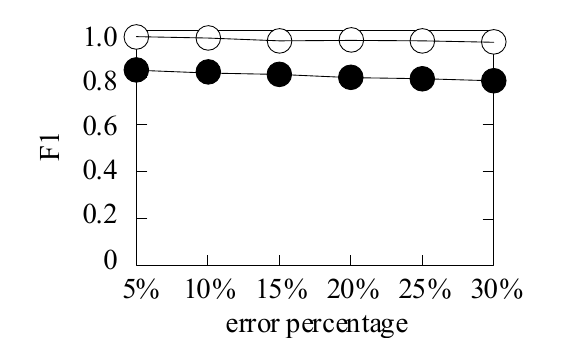}\label{fig:e-a-hai}
  }
  \subfigure[\emph{CAR}]{
   \includegraphics[width=1.6in]{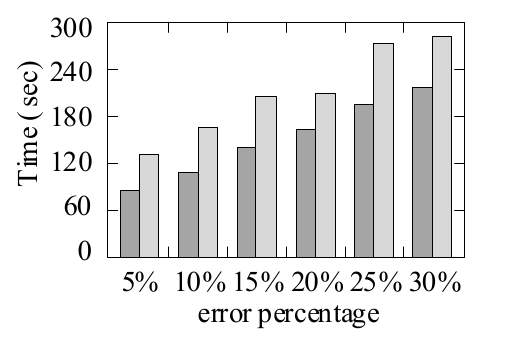}\label{fig:e-t-car}
  }%\hspace*{-5mm}
\subfigure[\emph{HAI}]{
   \includegraphics[width=1.6in]{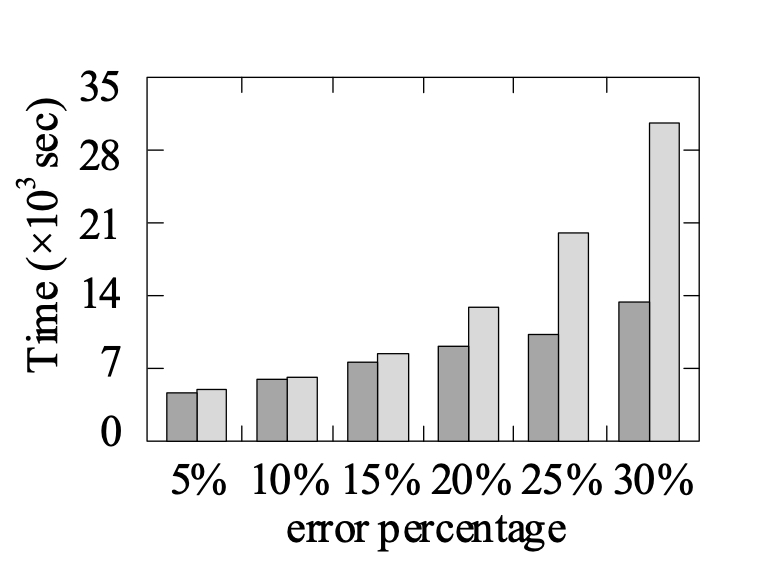}\label{fig:e-t-hai}
  }
\vspace*{-2mm}
\caption{Effect of error percentage on comparison evaluation}
\label{fig:e-comparison}
\vspace*{-2mm}
\end{figure*}

In the experiments, we use two real-world datasets, i.e., \emph{HAI} and \emph{CAR}, and one synthetic dataset, i.e., \emph{TPC-H}.

\textbf{HAI\footnote{\footnotesize https://data.medicare.gov/data/hospital-compare}}  is a real dataset that provides information about healthcare associated  infections occurred in hospitals. It contains 231,265 tuples.

\textbf{CAR\footnote{\footnotesize https://www.cars.com/}} contains the used vehicle information, including \emph{model, make, type, year, condition, wheelDrive, doors,} and  \emph{engine} attributes. It consists of 30,760 tuples.

\textbf{TPC-H\footnote{\footnotesize http://www.tpc.org/tpch/}} is a benchmark for performance metrics for systems operating. The two largest tables including \emph{lineitem} and \emph{customer} tables are utilized to create a synthetic dataset. It contains 6,001,115 tuples.

Table \ref{table:Dataset Rules} summarizes integrity constraints of each dataset used in our experiments, and they are given by domain experts.
We add errors randomly, including \emph{typos} and \emph{replacement errors}, on attributes related to integrity constraints shown in Table \ref{table:Dataset Rules} for each dataset. Specifically, we randomly delete any letter of an attribute value to construct a typo. For a replacement error, we replace a value with another value from the same domain. For each dataset, we generate $5\%$ error rate by default, including a half fraction of typos and another half fraction of replacement errors. It is necessary to mention that, enterprises typically find data error rates are approximately $5\%$ \cite{fan2012foundations}, and the reported error rates are no more than $30\%$ in many case studies \cite{redman1998impact}.
Note that, the error rate is defined as the number of erroneous values to the number of total attribute values. In addition, we use the Levenshtein distance as the distance metric, unless otherwise stated.

We utilize F1-score to evaluate the accuracy of data cleaning methods. Specifically, it is defined as
\begin{equation}
\label{equation:evaluation}
\text{F1-score} = \frac{2\times precision \times recall}{precision + recall}
\end{equation}
where \emph{precision} is equal to the ratio of correctly repaired attribute values to the total number of updated attribute values, and \emph{recall} equals the ratio of correctly repaired attribute values to the total number of erroneous values.
In addition, unless otherwise stated, the  experiments  were conducted on a Dell PowerEdge T620 with one Intel(R) Xeon(R) E5-2620 v2 2.10GHz processors (6 physical cores and 12 CPU threads) and 188GB RAM.

%\begin{figure}[t]
%\centering
%\hspace*{-3mm}
%\subfigure[F1-score]{
%   \includegraphics[height=1.2in]{fig/CAR-similarity.eps}
%  }\hspace*{-8mm}
%\subfigure[runtime]{
%   \includegraphics[height=1.2in]{fig/HAI-similarity.eps}
%  }
%\vspace*{-4mm}
%\caption{The impact of different distance algorithms on F1-score for both CAR and HAI datasets}
%\label{fig:similar-centre}
%\vspace*{-5.5mm}
%\end{figure}

\subsection{Comparisons with HoloClean}
\label{sec:compared_with_holoClean}

In this section, we verify the performance of \textsf{MLNClean} and HoloClean.
Since HoloClean adopts external modules for error detection and it can only fix errors caught by the error detection phase, we set the detection accuracy of HoloClean as 100\% for an absolutely fair evaluation, which helps avoid the effect of the detection accuracy on the subsequent error repairing phase.
%we use ground truth file as the input dataset to learn the parameters of the statistical model, i.e., the detection accuracy is 100\%.

\textbf{Effect of error percentage.} We change the error percentage from 5\% to 30\%, and report the corresponding experimental results in Figure \ref{fig:e-comparison}. As observed from Figure \ref{fig:e-a-car} and Figure \ref{fig:e-a-hai}, for both \textsf{MLNClean} and HoloClean, the accuracy decreases slightly as the error percentage increases. For \textsf{MLNClean}, there are two reasons for the decline. The first reason is that, with the increase of error percentage, \textsf{AGP} is prone to wrongly treat more normal groups as abnormal ones, and the following cleaning steps are subject to the chain reaction of \textsf{AGP}, resulting in degraded accuracy.
The second reason originates in statistical characteristics of \textsf{RSC}, which employs the reliability score based on Markov weight learning. The larger error percentage leads to the less reliability of the learned weight. On the other hand, HoloClean separates the whole dataset into noisy and clean parts. It uses clean values which are picked by error detection methods to learn the statistical model parameters. Then, it employs the trained model to infer the probability of each noisy value. When error rate increases, the statistical difference between the noisy and clean parts enlarges, which incurs the unsuitable parameters for the inference of noisy values.
The results also show that, \textsf{MLNClean} has much higher accuracy than  HoloClean for all cases, which reflects the superiority of the two-stage cleaning of \textsf{MLNClean}. This is because, when erroneous values become more and more, HoloClean relying solely on probabilistic reasoning is becoming weaker. However, \textsf{MLNClean} considering both statistical characteristics and the principle of minimality is much stronger, thereby it has better performance.

%\begin{figure}[t]
%\centering
%\vspace*{-2mm}\hspace*{3mm}
%\vspace*{-3mm}\includegraphics{fig/time-icon.eps}
%\hspace*{-4mm}
%\subfigure[CAR]{
%   \includegraphics[width=1.6in]{fig/CAR-percentage-time.eps}
%  }\hspace*{-5mm}
%\subfigure[HAI]{
%   \includegraphics[width=1.6in]{fig/HAI-percentage-time.eps}
%  }
%\vspace*{-2mm}
%\caption{Runtime analysis of different total error percentage in both CAR and HAI datasets.}
%\label{fig:runtime-centre-percent}
%\vspace*{-4mm}
%\end{figure}

As shown in Figure \ref{fig:e-t-car} and Figure \ref{fig:e-t-hai}, in terms of the execution time, we can observe that the time cost increases when the error percentage is growing for both \textsf{MLNClean} and HoloClean.
Note that, the overall runtime of \textsf{MLNClean} includes both error detection time and repairing time, but the total runtime of HoloClean only involves the error repairing time due to its property.
%Note that, for the detailed analysis, the runtime of \textsf{MLNClean} is reported in two parts, i.e., the Markov weight learning time and the remaining cleaning time other than the weight learning (\textsf{MLNClean} cleaning time for short), which includes both the error detection and repairing time.
%In contrast, due to the property of HoloClean, the total runtime of HoloClean only involves the error repairing time.
For \textsf{MLNClean}, the growth of the total time cost mainly results from the Markov weight learning, which occupies almost 95\% of the total time. Specifically, the increase of error percentage makes it more difficult to determine whether a value is clean or erroneous, thus leading to the slower convergence of Markov weight learning.
%{\color{blue}(ii) The more the errors, the more $\gamma$s a group contains. Hence, it takes longer time to execute \textsf{RSC}.
%(iii) The more the errors, the more likely the conflicts are to occur. The runtime of the conflict resolution is linearly related to the number of $\gamma$s containing conflicts.}
For HoloClean, the runtime is mainly determined by the compile and repair phase. In particular, for more errors, the candidate set of each value turns larger, incurring more overhead.
%the runtime increases since the runtime of compile and repair is positively related to the size of the domain.

Furthermore, \textsf{MLNClean} is consistently faster than HoloClean, even though \textsf{MLNClean} deals with both error detection and repairing stages and HoloClean only focuses on the error repairing.
The superiority of \textsf{MLNClean} comes from the cleaning scheme of \textsf{MLNClean}.
In \textsf{MLNClean}, the smallest unit of data cleaning is a piece of data, i.e., $\gamma$, which contains multiple attribute values. Thus, it is able to clean multiple values per time. However, in HoloClean, the minimum cleaning unit is a single attribute value, namely, it only cleans one value per time. Hence, it needs longer time for HoloClean to clean all errors.

\textbf{Effect of error type ratio.}
In order to investigate the effect of different error types on the performance of \textsf{MLNClean} and HoloClean, we vary the error type ratio, and set the total error percentage as 5\% by default.
We consider two error types, i.e., \emph{replacement errors} and \emph{typos}.
Specifically, we change the proportion of replacement errors to the $5\%$ total errors, denoted as $R_{ret}$, from 0 to 100\%. %$R$ represent replacement error and $T$ represent typos.
In particular, $R_{ret}$ being zero means that there is no replacement error in the 5\% total errors, namely, all the 5\% total errors are typos. $R_{ret}$ being 100\% indicates that all the 5\% total errors are replacement errors.
%Assuming total errors account for 100, 0/100 means there are only typos in dataset while 100/0 means there exists only replacement errors. We change 25 error number in each experiment from 0/100 to 100/0. (ii) change total error percentage from 5\% to 30\%.

\begin{figure}[t]
\centering
%\hspace*{-3mm}
\includegraphics[width=0.33\textwidth]{fig/compare-icon.pdf}\\
\vspace*{-2mm}
\hspace*{-4mm}
\subfigure[\emph{CAR}]{
   \includegraphics[height=1.1in]{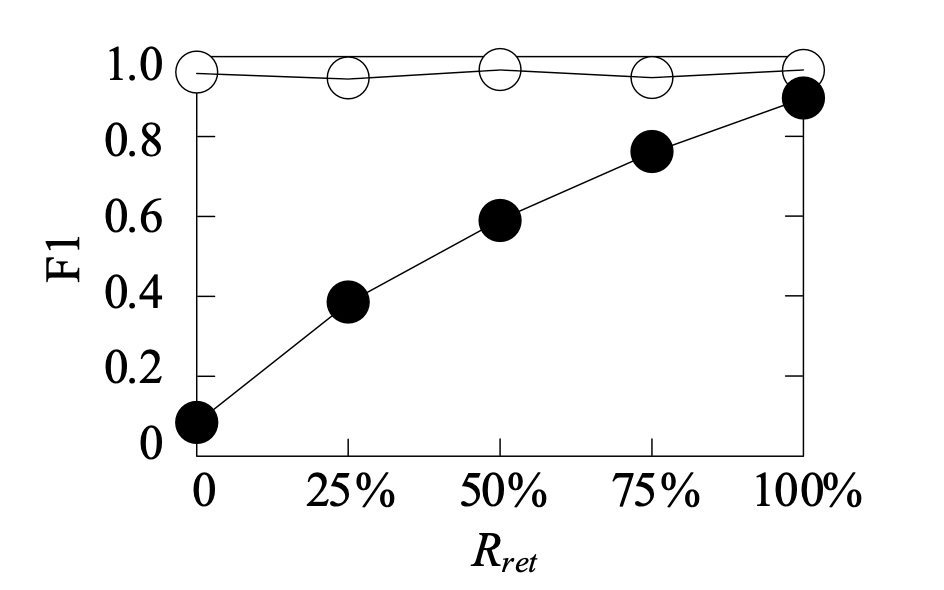}
  }%\hspace*{-8mm}
\subfigure[\emph{HAI}]{
   \includegraphics[height=1.1in]{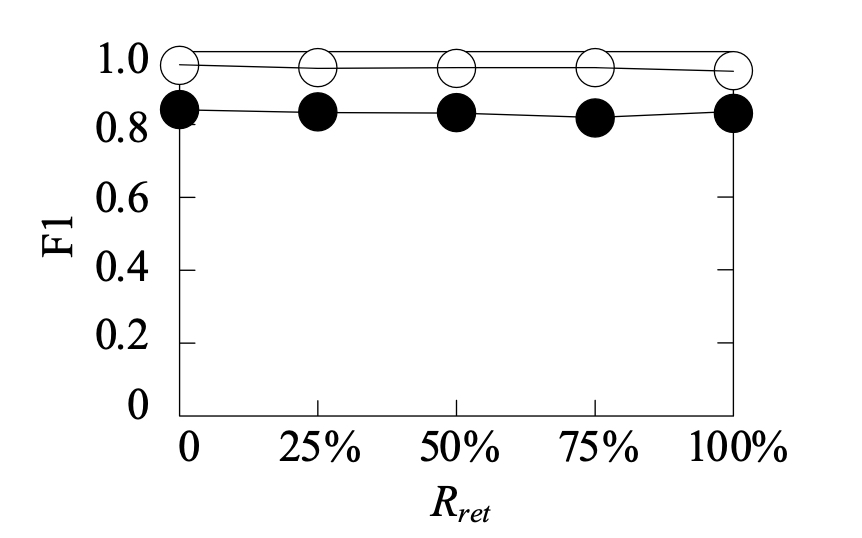}
  }
\vspace*{-4mm}
\caption{Effect of error type ratio on comparison evaluation}
\label{fig:f1-centre-ratio}
%\vspace*{-2mm}
\end{figure}

The corresponding experimental results are plotted in Figure \ref{fig:f1-centre-ratio}.
We can observe that, HoloClean is rather sensitive to the error type ratio $R_{ret}$  on the \emph{CAR} dataset. In contrast, the performance of \textsf{MLNClean} is stable on both datasets.
 %when varying error type ratio $R_{ret}$.
The reason behind mainly comes from two aspects. From the method aspect, as explained previously, HoloClean trains the model using the clean part, and infers values in the dirty part using the trained model to correct errors. Following the generation methods, the replacement errors are incorrect values from the corresponding same domain, and thus, they exist in both clean and dirty data parts.
In contrast, typos are absent from the clean part, which leads the trained model to be weak for many typos.
Thus, HoloClean is supposed to be sensitive to typos. From the dataset aspect, \emph{CAR} is rather sparse while \emph{HAI} is relatively much dense. Hence, HoloClean is much sensitive to $R_{ret}$ over \emph{CAR} than that over \emph{HAI}.
The F1-score of HoloClean in \emph{CAR} shows a growing trend when varying $R_{ret}$ from 0 to 100\%. Especially when there are only typos errors in the data set, the cleaning result is the worst. On the other hand, \textsf{MLNClean} fully considers both types of errors via the two-stage cleaning strategy. Thus, it is much stable with the change of $R_{ret}$. It further confirms the superiority of \textsf{MLNClean}. In addition, it is necessary to mention that, the execution time is not very sensitive to the error type. Hence, we omit the related description due to the space constraint.

\subsection{Results on MLNClean}

In this section, we study the effect of different parameters (i.e., the value of threshold in \textsf{AGP} strategy, the total error percentage, and the distance metric) on the performance of \textsf{MLNClean}.
Especially, for in-depth investigation, we also explore the effect of parameters on the three components of \textsf{MLNClean}, including \textsf{AGP} strategy, \textsf{RSC} method, and \textsf{FSCR} strategy, each of which has an impact on the data cleaning accuracy of \textsf{MLNClean}. In particular, in order to appropriately measure the accuracy of each component, we introduce a series of metrics for them.

For \textsf{AGP} strategy, we define \emph{Precision-A}  as the fraction of correctly merged abnormal groups over the total number of detected abnormal groups, and \emph{Recall-A} as the fraction of correctly merged abnormal groups over the total number of real abnormal groups. For \textsf{RSC} method, \emph{Precision-R} is defined as the ratio of correctly repaired $\gamma$s to the total number of repaired $\gamma$s, and \emph{Recall-R} is equal to the ratio of correctly repaired $\gamma$s to the number of $\gamma$s which contain errors. In addition, for \textsf{FSCR} strategy, \emph{Precision-F} (resp. \emph{Recall-F}) corresponds to the fraction of correctly repaired attribute values by \textsf{FSCR} over the number of erroneous attribute values that include detected conflicts (resp. the total number of erroneous attribute values).

%Then, we evaluate the performance of MLNClean comparing with HoloClean by reporting F1-score and runtime using HAI and CAR datasets.

%In this subsection, we study the effect of threshold on each stage of MLNClean.

\subsubsection{Effect of Threshold}
\begin{figure}[t]
\centering
%\vspace*{-6mm}
\includegraphics[width=0.32\textwidth]{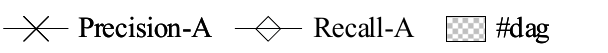}\\
%\vspace*{-1mm}
\hspace*{-4mm}
\subfigure[\emph{CAR}]{
   \includegraphics[height=1.1in]{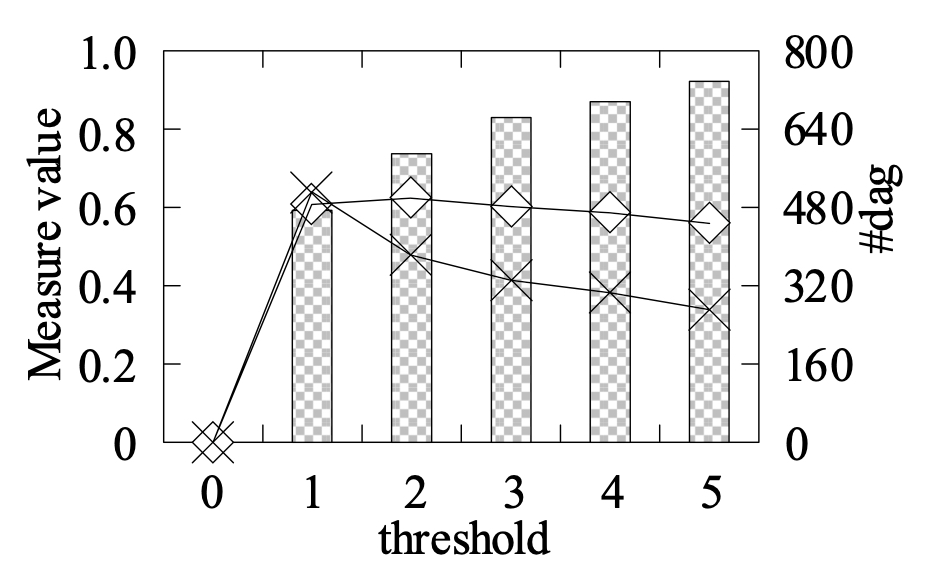}
  }\hspace*{-3mm}
\subfigure[\emph{HAI}]{
   \includegraphics[height=1.1in]{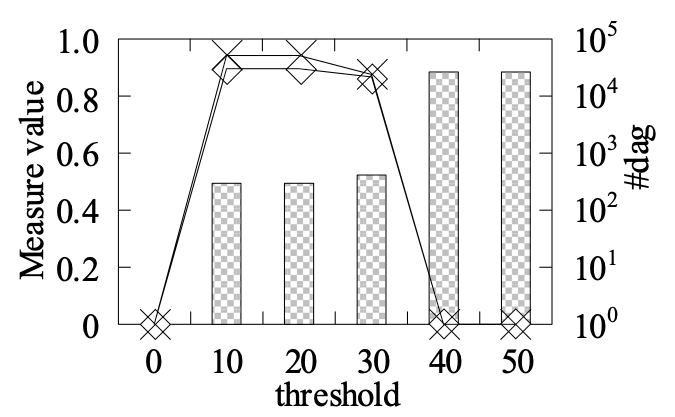}
  }
\vspace*{-4mm}
\caption{The performance of \textsf{AGP} vs. the value of $\tau$}
\label{fig:abnmg-threshold}
\vspace*{-2mm}
\end{figure}

%since the threshold is a direct factor in determining whether a group is abnormal
\textbf{Effect on the performance of \textsf{AGP}}.
First, we evaluate the performance of abnormal group process (i.e., \textsf{AGP}) strategy when varying the value of threshold $\tau$.
The corresponding results are shown in Figure \ref{fig:abnmg-threshold}.
Note that, we also report the total size of pieces of data (i.e., $\gamma$s) within detected abnormal groups by \textsf{AGP}  under different thresholds. For simplicity, we call it the number of detected abnormal $\gamma$s (i.e., \#dag for short).

The first observation is that, the accuracy of \textsf{AGP} (both precision and recall) first ascends and then drops as the value of $\tau$ grows over both datasets. \textsf{AGP} achieves the highest accuracy for \emph{CAR} dataset at $\tau$ being 1. For \emph{HAI} dataset, it achieves the highest accuracy when $\tau$ gets 10.
%{\color{blue}Furthermore, we find that the optimal accuracy on \emph{CAR} is not as good as that on \emph{HAI}. The reason is that, more replacement errors are added into the reason part in \emph{CAR}, which causes $\gamma$s containing these errors to form abnormal groups rather than normal groups.}
The second observation is that, when $\tau$ is zero, the accuracy is near to zero on both  datasets.
The reason is that no group is treated as abnormal in this case. It corresponds to the lowest size of abnormal $\gamma$s (i.e., \#dag depicted in diagrams). In contrast, when the value of $\tau$ exceeds the optimal value, the accuracy deteriorates, while the corresponding size of abnormal $\gamma$s increases over both datasets.
This is because, more normal groups are detected as abnormal groups.
In particular, there is an extreme situation that the accuracy sharply drops to zero, while the corresponding size of abnormal $\gamma$s grows significantly when $\tau$ is larger than 30 on \emph{HAI} dataset.
%This phenomenon is attributed to
The reason behind is that, the vast majority of normal groups are wrongly detected as abnormal ones.
%The results show that a appropriate threshold is critical for abnormality merging.

\begin{figure}[t]
\centering
%\vspace*{-2mm}
\includegraphics[width=0.24\textwidth]{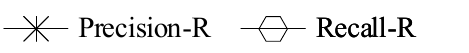}\\
\vspace*{-1.8mm}
%\hspace*{-2mm}%
\subfigure[\emph{CAR}]{
   \includegraphics[height=1.1in]{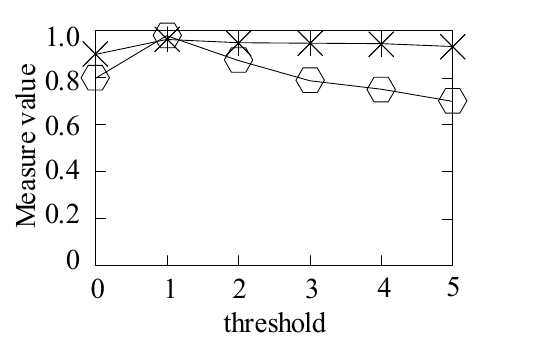}
  }%\hspace*{-1.5mm}
\subfigure[\emph{HAI}]{
   \includegraphics[height=1.1in]{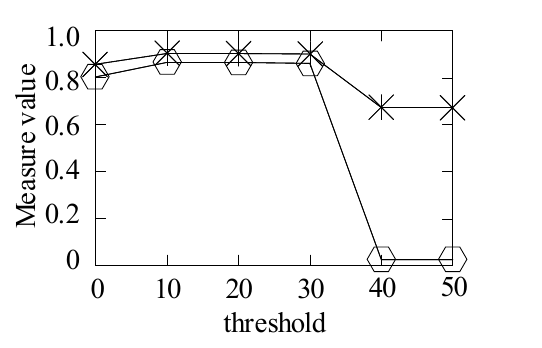}
  }
\vspace*{-4mm}
\caption{The performance of \textsf{RSC} vs. the value of $\tau$}
\label{fig:rs-threshold}
\vspace*{-2mm}
\end{figure}

\textbf{Effect on the performance of \textsf{RSC}}.
Then, we investigate the impact of the threshold on the accuracy of the reliability score based cleaning strategy (i.e., \textsf{RSC}).
 %by \emph{relscore precision} defined as the fraction of correctly repaired $\gamma$s over the total number of repaired $\gamma$s and \emph{relscore recall} defined as the fraction of correctly repaired $\gamma$s over the number of $\gamma$s which contain errors.
As shown in Figure \ref{fig:rs-threshold}, an appropriate value of threshold (i.e., $\tau$ being 1 on \emph{CAR} and $\tau$ being 10 on \emph{HAI}) contributes to the higher accuracy of \textsf{RSC}.
%optimal repair result with high relscore precision and relscore recall.
Nonetheless, when the value of $\tau$ deviates from the optimal value, the accuracy gets worse.
The reason is that, the further the $\tau$ is to the optimal value, the more the groups are processed wrongly by \textsf{AGP}, and thus, the less the pieces of data in groups are correctly repaired by \textsf{RSC}.
Besides, the precision of \textsf{RSC} (w.r.t. \emph{Precision-R}) remains higher than the recall of \textsf{RSC} (w.r.t. \emph{Recall-R}). This is because, when more groups are wrongly processed by \textsf{AGP} in previous step, \textsf{RSC} executed within each group is not able to repair more errors caused by \textsf{AGP}, resulting in the lower recall.
%it makes  few $\gamma$s are over-cleaned (i.e., few correct $\gamma$s are modified to errors).
There is such an extreme case that the recall sharply drops nearly to zero when $\tau$ is larger than 30 on \emph{HAI} dataset.
%
%this is because large number of regular groups are wrongly detected as abnormal groups in the previous phase (i.e., abnormality merging), lacking of sufficient regular groups as benchmarks to add abnormalities to the corresponding regular groups makes it unable to perform the merging operation, thus a large number of errors are ignored to repair in cleaning within group stage.

\begin{figure}[t]
\centering
%\vspace*{-1mm}
\includegraphics[width=0.24\textwidth]{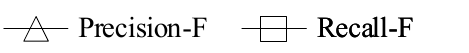}\\
\vspace*{-0.5mm}
%\hspace*{-4mm}
\subfigure[\emph{CAR}]{
   \includegraphics[height=1.1in]{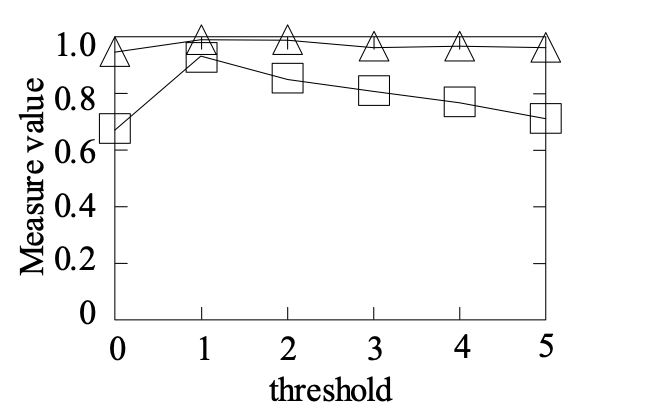}
  }%\hspace*{-3mm}
\subfigure[\emph{HAI}]{
   \includegraphics[height=1.1in]{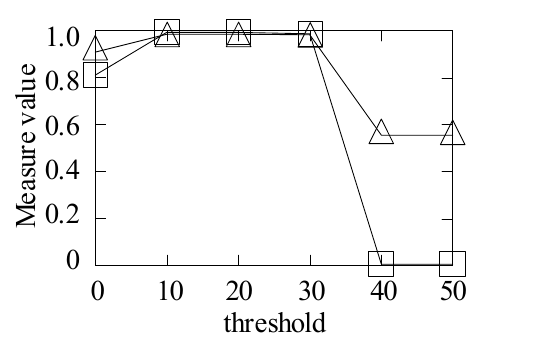}
  }
\vspace*{-4mm}
\caption{The performance of \textsf{FSCR} vs. the value of  $\tau$}
\label{fig:cf-threshold}
\vspace*{-2mm}
\end{figure}

\textbf{Effect on the performance of \textsf{FSCR}}. Next, we study the performance of our presented conflict resolution method (i.e., \textsf{FSCR}) when varying the value of threshold.
 %by \emph{conflres precision} and \emph{conflres recall}. The conflres precision is defined as the fraction of correctly repaired attribute values over the number of error attribute values that contains detected conflicts. And the conflres recall is defined as the fraction of correctly repaired attributes over the total number of error attribute values.
The corresponding results are depicted in Figure \ref{fig:cf-threshold}.
As expected, the appropriate value of threshold (i.e., $\tau$ being 1 on \emph{CAR} and  $\tau$ being 10 on \emph{HAI}) contributes to the optimal accuracy.
%While inappropriate threshold settings have a negative impact on the conflict resolution.
Moreover, we can find that, the precision maintains high value  when the value of $\tau$ deviates from the optimal value.
According to the definition of precision, it means that few detected conflicts are wrongly repaired.
%reflects that few clean attributes are over-cleaned. correctly repaired attribute values over the number of error attribute values that contains detected conflicts
%In particular, the drop of CAR in conflres precision is around $4\%$ when $\tau \in [0,5]$, the drop of HAI in conflres precision is around $7\%$ when $\tau \in [0,30]$.
Besides, the lower recall than precision indicates that, some errors have not been detected by \textsf{FSCR}.
The recall sharply drops, even near to zero for \emph{HAI} when $\tau$ is larger than 30.
It signifies that, there are more and more erroneous values that are not detected by \textsf{FSCR}, since those errors have not been correctly processed by \textsf{AGP} or \textsf{RSC} in previous phases.
%mThe reason behind is that, since large number of error attributes are not detected as conflicts because of the neglect repairing of them in the previous phase (i.e., cleaning within group) according to the reliability score.

\begin{figure}[t]
\centering
\vspace*{-3mm}
\includegraphics[width=0.32\textwidth]{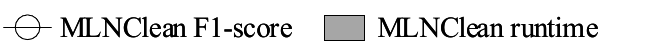}\\
\vspace*{-0.5mm}
\hspace*{-4mm}
\subfigure[\emph{CAR}]{
   \includegraphics[height=1.12in]{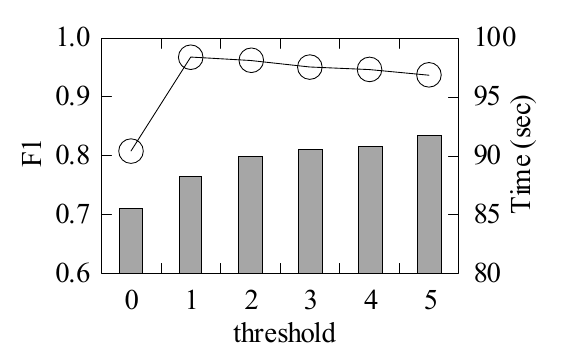}
  }\hspace*{-3mm}
\subfigure[\emph{HAI}]{
   \includegraphics[height=1.12in]{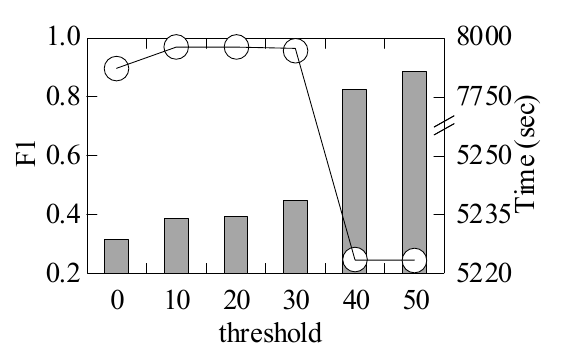}
  }
\vspace*{-4mm}
\caption{The performance of \textsf{MLNClean} vs. the value of $\tau$}
\label{fig:cf-mlnclean}
\vspace*{-2mm}
\end{figure}

\textbf{Effect on the performance of \textsf{MLNClean}.}
Last but not the least, we explore the effect of threshold on the overall framework \textsf{MLNClean}, and report the corresponding experimental results in Figure~\ref{fig:cf-mlnclean}.
It is hardly surprising that, \textsf{MLNClean} gets the highest accuracy  when $\tau$ is 1 on \emph{CAR} dataset (in which  F1 is 0.96), and when $\tau$ equals 10 on \emph{HAI} dataset (where F1 equals 0.98).
The deviation of the threshold value from the most appropriate value contributes to the descend of accuracy.
On the other  hand, the total execution time of \textsf{MLNClean} turns longer as the increasing value of threshold. This is because, the bigger the value of threshold, the larger the number of detected abnormal groups by \textsf{AGP}, and hence, it leads to the longer processing time. Note that, without loss of generality, we set $\tau$ as 1 on \emph{CAR} dataset (and as 10 on \emph{HAI} dataset) in the rest of experiments.

\subsubsection{Effect of Error Percentage}
\label{sec:effect_error_percentage_MLNClean}

%In this subsection, we evaluate the effect of error percentage on each cleaning stage of MLNClean by varing error percentage from 5\% to 30\% for the reason that enterprises typically find data error rates are approximately $5\%$ \cite{fan2012foundations} and the reported error rates are no more than $30\%$ in many case studies \cite{redman1998impact}.

\begin{figure}[t]
\centering
%\vspace*{-6mm}
\includegraphics[width=0.32\textwidth]{fig/threshold-icon.pdf}
%\vspace*{-1mm}
\hspace*{-5mm}
\subfigure[\emph{CAR}]{
   \includegraphics[height=1.1in]{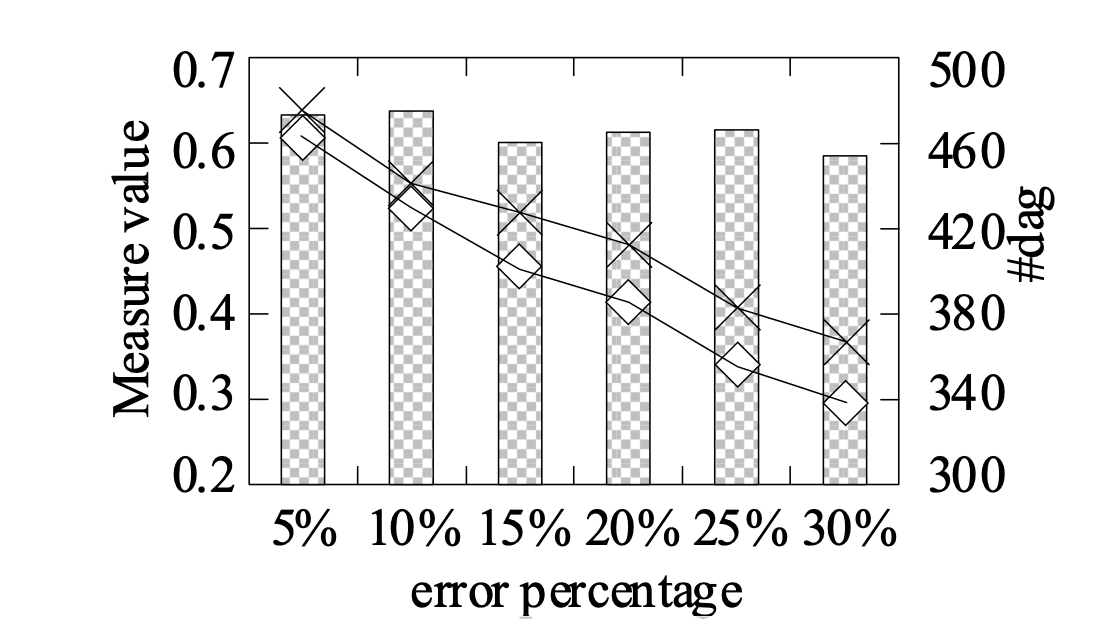}
  }\hspace*{-7mm}
\subfigure[\emph{HAI}]{
   \includegraphics[height=1.1in]{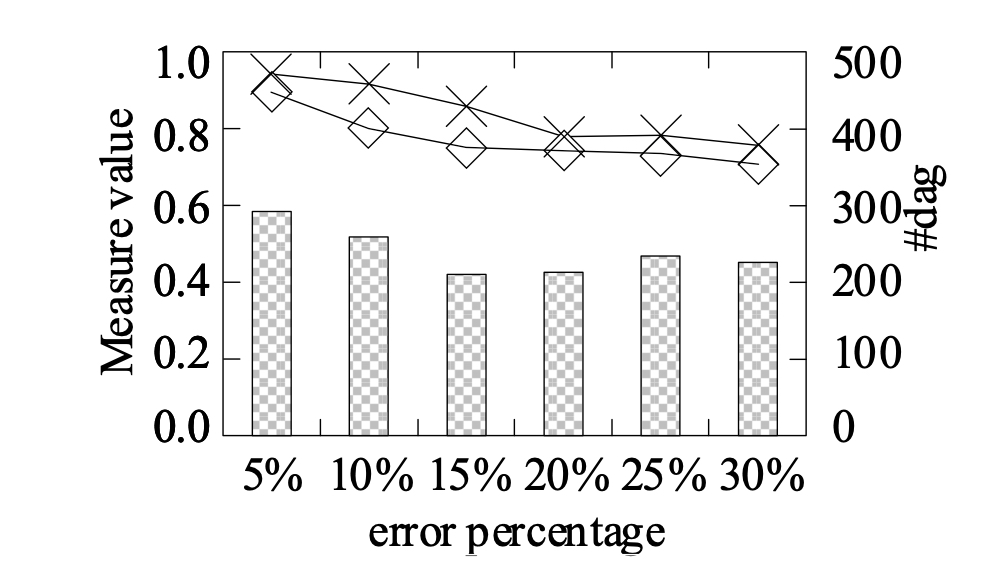}
  }
\vspace*{-4mm}
\caption{The performance of \textsf{AGP} vs. the error percentage}
\label{fig:am-errorrate}
%\vspace*{-2mm}
\end{figure}

\textbf{Effect on the performance of  \textsf{AGP}}. First, we verify the effect of error percentage on the performance of \textsf{AGP}. The corresponding results are shown in Figure \ref{fig:am-errorrate} with various error percentages.
It is observed that, as the growth of error percentage, the accuracy of \textsf{AGP} decreases.
This is because, the higher the error rate, the more the (abnormal) groups, and hence the less the $\gamma$s within one group.
Thus, when there are less and less $\gamma$s within a group with the increase of error rate, \textsf{AGP} easily tends to treat more and more normal groups wrongly as abnormal ones, for a fixed value of threshold on each dataset.
As a result, both the precision and recall of \textsf{AGP} gets lower according to their definitions.

%While when the error percentage increases, it easily tends to treat some normal groups wrongly as abnormal ones in the case that the size of those normal groups are smaller than the size of the abnormal groups.

\begin{figure}[t]
\centering
\vspace*{-2.3mm}
\includegraphics[width=0.24\textwidth]{fig/rs-threshold-icon.pdf}\\
\vspace*{-1mm}
\hspace*{-4mm}
\subfigure[\emph{CAR}]{
   \includegraphics[height=1.1in]{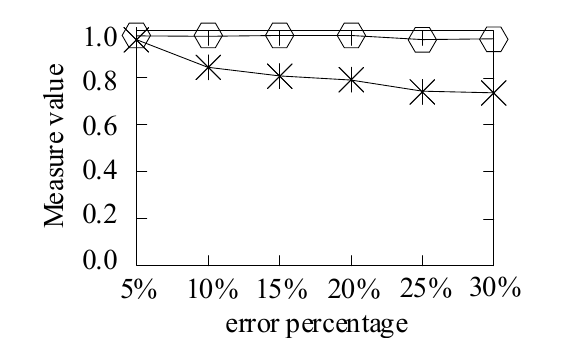}
  }\hspace*{-3mm}
\subfigure[\emph{HAI}]{
   \includegraphics[height=1.1in]{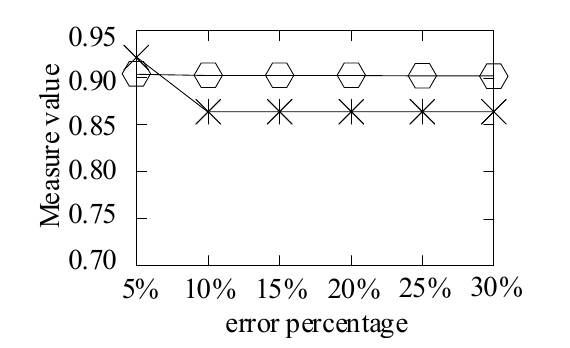}
  }
\vspace*{-4mm}
\caption{The performance of \textsf{RSC} vs. the error percentage}
\label{fig:rs-errorrate}
%\vspace*{-2mm}
\end{figure}

\textbf{Effect on the performance of  \textsf{RSC}}.
Then, we investigate the accuracy of \textsf{RSC} method when changing the error percentage, and report the corresponding results in Figure \ref{fig:rs-errorrate}.
We can observe that, both the precision and recall of \textsf{RSC} drop slightly with the increasing error rate.
There are two major reasons for the trends.
The first one is about the propagated influence of the decreasing accuracy of \textsf{AGP} in the previous step.
The second reason comes from the statistical characteristic of \textsf{RSC}, which employs a reliability score based on Markov weight learning. The larger the error rate, the less reliable the learned weights, and the lower the accuracy of \textsf{RSC}.
%{\color{blue}(i)for probability factor, the difference of probability between the correct $\gamma$s and error $\gamma$s within the same group is narrow down, even the probability of a correct $\gamma$ is lower than a error $\gamma$ in some cases; (ii) for distance factor, it is more likely that the number of error $\gamma$s is more than correct $\gamma$s in some groups, thus the error $\gamma$s gain larger distance according to the principle of minimality.}
On the other hand, we have to mention that, \textsf{RSC} is quite robust to the change of error rate. In particular, the drop of precision is around 10\%, and the drop of recall is around 1\%.
In addition, the recall is higher than the precision in most cases.
The reason is that, with the growth of error rate, the number of repaired $\gamma$s by \textsf{RSC} increases faster than the number of $\gamma$s containing errors, which  partly results from more wrongly processed groups of \textsf{AGP} (as explained earlier).
%For HAI dataset, the drop in relscore precision and relscore recall is around 0.1\%.

\begin{table}[!]
\centering\small
\vspace*{-2mm}
\setlength{\tabcolsep}{15pt}
\caption{F1-scores under Different Distance Metrics}
\begin{tabular}{|c|c|c|}
\hline
                              & \textbf{Levenshtein distance} & \textbf{Cosine distance}  \\ \hline
 \emph{CAR} & 0.968      & 0.730      \\ \hline
  \emph{HAI}    &0.970        & 0.947     \\ \hline
\end{tabular}
\label{table:f1-distance-metrics}
\vspace*{-3mm}
\end{table}

\begin{figure}[t]
\centering
\vspace*{-2mm}
\includegraphics[width=0.24\textwidth]{fig/cf-threshold-icon.pdf}\\
\vspace*{-1mm}
\hspace*{-4mm}
\subfigure[\emph{CAR}]{
   \includegraphics[height=1.1in]{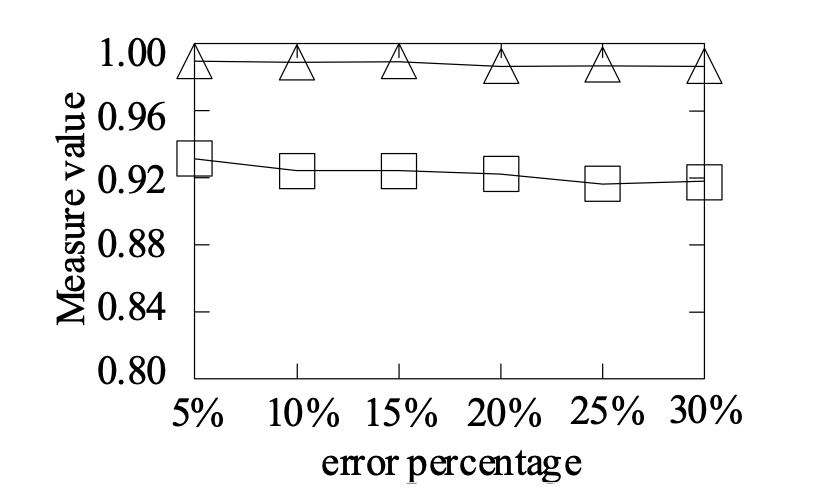}
  }\hspace*{-5mm}
\subfigure[\emph{HAI}]{
   \includegraphics[height=1.1in]{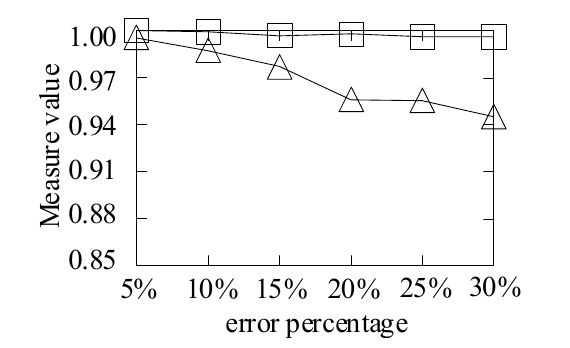}
  }
\vspace*{-6mm}
\caption{The performance of \textsf{FSCR} vs. the error percentage}
\label{fig:cf-errorrate}
\vspace*{-2mm}
\end{figure}

\textbf{Effect on the performance of \textsf{FSCR}}. We also study the impact of error percentage on the performance of \textsf{FSCR}.  The corresponding results are depicted in Figure \ref{fig:cf-errorrate}.
One can observe that, the accuracy has no significant fluctuation with the changing error percentage.
The values of both precision and  recall are always above 90\%, and the fluctuation  of them is within 6\%.
The high accuracy of \textsf{FSCR} reflects that, \textsf{FSCR} is indeed capable of cleaning out those errors which have not been correctly cleaned by \textsf{AGP} or \textsf{RSC} in previous stages.
Furthermore, it attributes the higher recall than precision to more wrongly detected conflicts by  \textsf{FSCR}, due to the relatively lower accuracy of \textsf{AGP} and \textsf{RSC} in previous stages.
In addition, it is worth pointing out that, we have analyzed the overall performance of \textsf{MLNClean} in terms of the experimental results in Figure~\ref{fig:e-comparison} when changing the error percentage. Thus, we omit the related description here due to the space constraint.

%\subsubsection{The Impact of Distance Metrics on MLNClean}
%\label{sec:distance_metric}
%In this section, we study the impact of different distance metrics on F1-score of MLNClean. In MLNClean, distance metrics integrates edit distance (ED) and cosine distance (CD) methods. We change the proportion of R/T to evaluate the pros and cons of these two distance metrics in MLNClean. As shown in Figure \ref{fig:similar-centre}, the result of Levenshtein distance is better than cosine distance overall. Cosine distance is more sensitive to different types of errors, it performs better when there are more replacement errors in dataset than in the case of more typos. Thus in the other experiments we use Levenshtein distance rather than cosine distance. In addition, the runtime is less affected by different distance metrics.

%\begin{figure}[t]
%\centering
%%\vspace*{-6mm}
%\includegraphics{fig/distance_metric_icon.eps}
%\hspace*{-4mm}
%\subfigure[CAR]{
%   \includegraphics[height=1.1in]{fig/distance_metric.eps}
%  }\hspace*{-3mm}
%\subfigure[HAI]{
%   \includegraphics[height=1.1in]{fig/distance_metric.eps}
%  }
%\vspace*{-4mm}
%\caption{Analyze the effect of changing the error percentage on the conflict resolution phase of MLNClean.}
%\label{fig:cf-errorrate}
%\vspace*{-2mm}
%\end{figure}

\subsubsection{Effect of Distance Metrics}

\begin{figure}[!]
\centering
%\vspace*{-1mm}
\includegraphics[width=0.32\textwidth]{fig/spark-icon.pdf}\\
\vspace*{-1mm}
\hspace*{-4mm}
\subfigure[\emph{HAI}]{
   \includegraphics[height=1.12in]{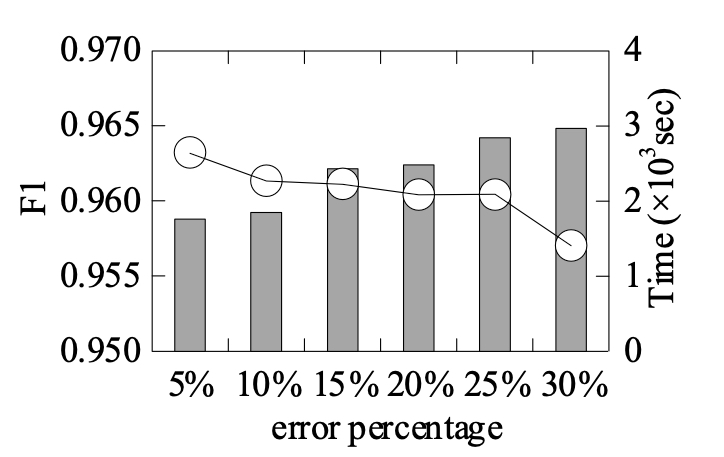}
  }\hspace*{-3mm}
\subfigure[\emph{TPC-H}]{
   \includegraphics[height=1.12in]{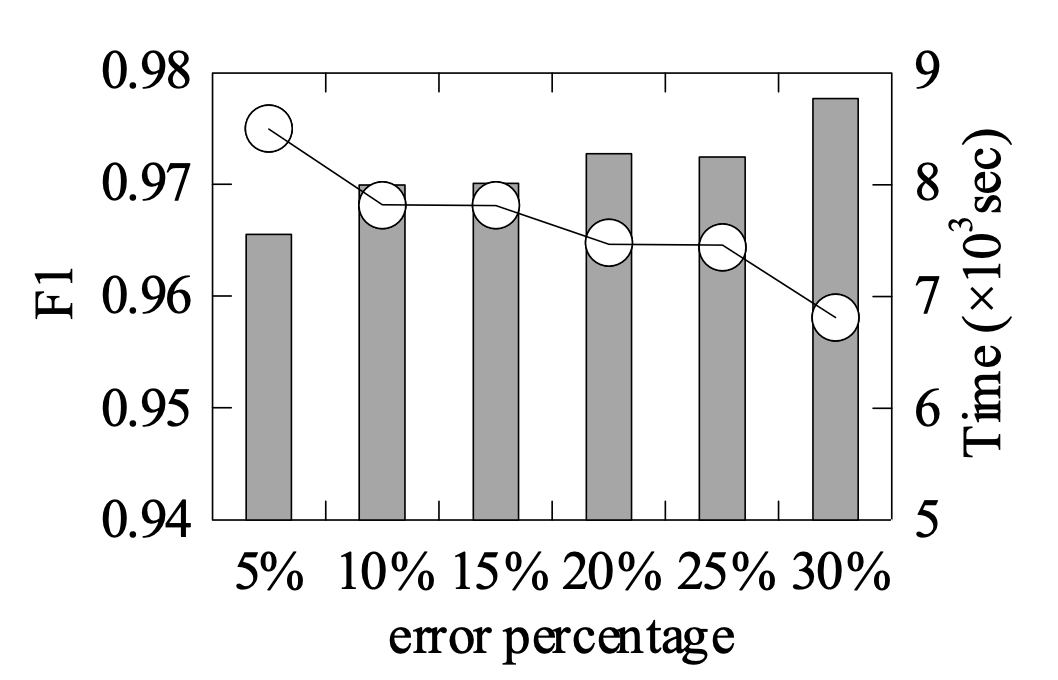}
  }
\vspace*{-4mm}
\caption{The performance of distributed \textsf{MLNClean}}
\label{fig:percent-spark}
\vspace*{-4mm}
\end{figure}

\begin{table}[!]
\centering
\small\setlength{\tabcolsep}{1pt}
\caption{Experiments under Different Numbers of Workers}
\begin{tabular}{|l|ccccc|}
\cline{1-6}
                           & \multicolumn{5}{c|}{ { \textbf{The number of workers}}}                                                           \\ \cline{2-6}
\multirow{-2}{*}{}         & \multicolumn{1}{c|}{2}     & \multicolumn{1}{c|}{4}     & \multicolumn{1}{c|}{6}     & \multicolumn{1}{c|}{8}     & \multicolumn{1}{c|}{10}   \\ \hline
%\textbf{Learning Time (s)} & \multicolumn{1}{c|}{50755} & \multicolumn{1}{c|}{27570} & \multicolumn{1}{c|}{16285} & \multicolumn{1}{c|}{11567} & \multicolumn{1}{c|}{7573} \\ \hline
\textbf{Total time of \textsf{MLNClean} (sec)}    & \multicolumn{1}{c|}{50,759} & \multicolumn{1}{c|}{27,574} & \multicolumn{1}{c|}{16,289} & \multicolumn{1}{c|}{11,572} & \multicolumn{1}{c|}{7,578} \\ \hline
\end{tabular}
\vspace*{-2mm}
\label{table:scalable_workers}
\end{table}

The distance metric plays an important role in \textsf{MLNClean} from two aspects. First, it is the basis of measuring the similarity of two groups, involving \textsf{AGP} strategy.  Second, it is an significant factor of computing the reliability score, which is employed by \textsf{RSC} method. Thus, we also evaluate the effect of different distance metrics, including the Levenshtein distance and  cosine distance, on the accuracy of \textsf{MLNClean}. As shown in Table \ref{table:f1-distance-metrics}, the accuracy of \textsf{MLNClean} using the Levenshtein distance is higher than that using cosine distance on both datasets. The reason is that, for the cosine distance, if the foremost few characters of a string are incorrectly spelled, the cosine distance from it to its similar string might be large.
Nevertheless, the Levenshtein distance just decides how many different characters between two strings, regardless of the positions of those characters. %does not consider the positions of different characters in strings, and it  to get the Levenshtein distance.
Thus, the Levenshtein distance is more suitable to deal with various error types.

%two main reasons as follows: (i) MLNClean provides a fine-grained partition methods to make a large Markov Network into multiple smaller ones. (ii) MLNClean performs a more efficient cleaning method than HoloClean.

%As shown in Figure \ref{fig:runtime-centre}(a) and Figure \ref{fig:runtime-centre}(b), when we vary total error percentage from $5\%$ to $30\%$, the runtime of both system is increasing on each stage. However, the runtime of HoloClean on each stage is more affected by total error percentage. Because

\subsection{Results on Distributed MLNClean}%¸ÄÁË×ÖÌå¸ñʽ£¬È¡Ïû\textsf{MLNClean},ÏԵò»Í³Ò»
\label{sec:distribute_exp}

%\begin{figure}[t]
%\centering
%\hspace*{3mm}
%\vspace*{-2mm}\includegraphics{fig/spark-icon.eps}
%\hspace*{-3mm}
%\subfigure[HAI]{
%   \includegraphics[height=1.15in]{fig/HAI-ratio-spark.eps}
%  }\hspace*{-4mm}
%\subfigure[TPC-H]{
%   \includegraphics[height=1.15in]{fig/TPCH-ratio-spark.eps}
%  }
%\vspace*{-4mm}
%\caption{Analyze the effect of changing the ratio of different error types on runtime and F1-score on Spark}
%\label{fig:ratio-spark}
%\vspace*{-4mm}
%\end{figure}

In this section, we evaluate the performance of our proposed distributed \textsf{MLNClean} version using larger \emph{HAI} and \emph{TPC-H} datasets. %We do not use CAR dataset since the dataset size of CAR is too small for distributed experiments. In addition,
This set of experiments was implemented on Spark 1.0.2, and was executed on a 11-node Dell cluster (1 master with 10 workers), each node has two Intel(R) Xeon(R) E5-2620 v3 2.4GHz processors (12 physical cores, 24 CPU threads) and 64GB RAM.

%In Figure \ref{fig:ratio-spark}, we verify the change of F1-score value as well as runtime when alter the proportion $R/T$. As shown in Figure \ref{fig:ratio-spark}(a) the result of HAI dataset show that the average $F_1=0.956$ and the average runtime$=1871$ seconds. And as shown in Figure \ref{fig:ratio-spark}(b) the result of TPC-H dataset show that the average $F_1=0.973$ and the average runtime$=7671$ seconds. In both HAI and TPC-H datasets, F1-score and runtime these two factors both perform evenly when the proportion $R/T$ changed. These experiments show that MLNClean can support different error types and achieve good cleaning results as well.

%\textbf{Runtime and F1-score Analysis.}
Figure \ref{fig:percent-spark} plots the corresponding results when varying the error percentage. % from $5\%$ to $30\%$, and report in
As expected, with the growth of error percentage, the execution time of \textsf{MLNClean} gets longer, and its accuracy gets lower on both \emph{HAI} and \emph{TPC-H} datasets. The reason behind is similar to that analyzed in Section \ref{sec:compared_with_holoClean}.
We would like to point out that, when the error percentage increases from 5\% to 30\%, the accuracy of \textsf{MLNClean} is always above 95\% for all cases, and the drop of accuracy is about less than 3\% over both datasets. Consequently, \textsf{MLNClean} maintains good robustness on Spark platform.

In addition, we change the number of workers from 2 to 10 on \emph{TPC-H} dataset, and present the corresponding results in Table \ref{table:scalable_workers}. One can observe that, the time cost drops as the number of workers grows, while the accuracy has very slight fluctuation. When the number of workers changes from 2 to 10, the efficiency has about 6.7 times speedup. %The reason is that, the more the workers, the smaller dataset in each worker is.

%\begin{figure}[t]
%\centering
%\includegraphics[height=1.15in]{fig/TPCH-workers.eps}
%\vspace*{-3mm}
%\caption{Scalability experiment by changing the number of workers}
%\label{fig:scalable_workers}
%\vspace*{-6mm}
%\end{figure}

\section{Conclusions}
\label{sec:conclude}

In this paper, we propose  a novel hybrid data cleaning framework  \textsf{MLNClean} on top of Markov logic networks (MLNs).
It combines the advantages of quantitative methods and qualitative ones, and is capable of cleaning both schema-level and instance-level errors. With the help of an effective two-layer \emph{MLN index}, \textsf{MLNClean} consists of two major cleaning stages, i.e., first cleaning multiple data versions independently and then deriving the final unified clean data from multi-version data. 
%The MLN index has a set of \emph{blocks} in the first layer and a set of \emph{groups} within each block in the second layer.
In the first cleaning stage, an \textsf{AGP} strategy is presented to process abnormal groups (built on the MLN index).
Based on a new concept of \emph{reliability score}, an \textsf{RSC} method is developed to clean data within each group.
Moreover, in the second cleaning stage, with a newly defined concept of \emph{fusion score}, an \textsf{FSCR} algorithm  is proposed to eliminate conflicts when unifying multiple data versions. 
% MLNClean, a two-stage data cleaning framework, combining qualitative and quantitative techniques to deal with both error detecting and repairing stages. The first stage cleans errors that violate integrity constraints, which gives the multi-version local optimal cleaning results for errors. The second stage unifies the multi-version cleaning results and cleans errors that are not correctly cleaned in the first stage. We introduce an optimization named MLN index structure for efficiency enhancement. %Moreover, we deploy MLNClean on the distributed Spark platform with an effective data partition strategy.
Extensive experimental results on both real and synthetic datasets demonstrate the superiority of \textsf{MLNClean} to the state-of-the-art approach in terms of both accuracy and efficiency. In the future, we intend to establish more sophisticated strategies to process abnormal groups, since the performance of this step significantly affects the overall performance of \textsf{MLNClean}.
%Also, we would like to gene
%First, since the Markov weight learning is costly, how to accelerate this learning is a direction worth studying. Second, instead of using \textsf{AGP} with an appropriate value of threshold, it is desirable to ,

%\vspace*{0.1in}

\section*{Acknowledgments}
This work was supported in part by the National Key R\&D Program of China under Grant No. 2018YFB1004003, the 973 Program under Grant No. 2015CB352502, the NSFC under Grant No. 61522208, the NSFC-Zhejiang Joint Fund under Grant No. U1609217, and the ZJU-Hikvision Joint Project. Both Yunjun Gao and Xiaoye Miao are the corresponding authors of the work.

\bibliographystyle{ieeetr}
\bibliography{MLNClean}

\begin{thebibliography}{10}

\bibitem{eckerson2002data}
W.~W. Eckerson, ``Data quality and the bottom line: Achieving business success
  through a commitment to high quality data,'' {\em The Data Warehousing
  Institute}, pp.~1--36, 2002.

\bibitem{chu2016data}
X.~Chu, I.~F. Ilyas, S.~Krishnan, and J.~Wang, ``Data cleaning: Overview and
  emerging challenges,'' in {\em SIGMOD}, pp.~2201--2206, 2016.

\bibitem{rahm2000data}
E.~Rahm and H.~H. Do, ``Data cleaning: Problems and current approaches,'' {\em
  IEEE Data Eng. Bull.}, vol.~23, no.~4, pp.~3--13, 2000.

\bibitem{abedjan2015temporal}
Z.~Abedjan, C.~G. Akcora, M.~Ouzzani, P.~Papotti, and M.~Stonebraker,
  ``Temporal rules discovery for web data cleaning,'' {\em {PVLDB}}, vol.~9,
  no.~4, pp.~336--347, 2015.

\bibitem{bertossi2005complexity}
L.~Bertossi, L.~Bravo, E.~Franconi, and A.~Lopatenko, ``Complexity and
  approximation of fixing numerical attributes in databases under integrity
  constraints,'' in {\em International Workshop on Database Programming
  Languages}, pp.~262--278, Springer, 2005.

\bibitem{beskales2010sampling}
G.~Beskales, I.~F. Ilyas, and L.~Golab, ``Sampling the repairs of functional
  dependency violations under hard constraints,'' {\em {PVLDB}}, vol.~3, no.~1,
  pp.~197--207, 2010.

\bibitem{beskales2013relative}
G.~Beskales, I.~F. Ilyas, L.~Golab, and A.~Galiullin, ``On the relative trust
  between inconsistent data and inaccurate constraints,'' in {\em ICDE},
  pp.~541--552, 2013.

\bibitem{bohannon2007conditional}
P.~Bohannon, W.~Fan, F.~Geerts, X.~Jia, and A.~Kementsietsidis, ``Conditional
  functional dependencies for data cleaning,'' in {\em ICDE}, pp.~746--755,
  2007.

\bibitem{bohannon2005cost}
P.~Bohannon, M.~Flaster, W.~Fan, and R.~Rastogi, ``A cost-based model and
  effective heuristic for repairing constraints by value modification,'' in
  {\em SIGMOD}, pp.~143--154, 2005.

\bibitem{chu2013holistic}
X.~Chu, I.~F. Ilyas, and P.~Papotti, ``Holistic data cleaning: Putting
  violations into context,'' in {\em ICDE}, pp.~458--469, 2013.

\bibitem{cong2007improving}
G.~Cong, W.~Fan, F.~Geerts, X.~Jia, and S.~Ma, ``Improving data quality:
  Consistency and accuracy,'' in {\em PVLDB}, pp.~315--326, 2007.

\bibitem{dallachiesa2013nadeef}
M.~Dallachiesa, A.~Ebaid, A.~Eldawy, A.~K. Elmagarmid, I.~F. Ilyas, M.~Ouzzani,
  and N.~Tang, ``{NADEEF}: {A} commodity data cleaning system,'' in {\em
  SIGMOD}, pp.~541--552, 2013.

\bibitem{fan2008conditional}
W.~Fan, F.~Geerts, X.~Jia, and A.~Kementsietsidis, ``Conditional functional
  dependencies for capturing data inconsistencies,'' {\em {ACM} Trans. Database
  Syst.}, vol.~33, no.~2, pp.~6:1--6:48, 2008.

\bibitem{geerts2013llunatic}
F.~Geerts, G.~Mecca, P.~Papotti, and D.~Santoro, ``The {LLUNATIC} data-cleaning
  framework,'' {\em {PVLDB}}, vol.~6, no.~9, pp.~625--636, 2013.

\bibitem{khayyat2015bigdansing}
Z.~Khayyat, I.~F. Ilyas, A.~Jindal, S.~Madden, M.~Ouzzani, P.~Papotti,
  J.~Quian{\'{e}}{-}Ruiz, N.~Tang, and S.~Yin, ``Big{D}ansing: {A} system for
  big data cleansing,'' in {\em SIGMOD}, pp.~1215--1230, 2015.

\bibitem{kolahi2009approximating}
S.~Kolahi and L.~V.~S. Lakshmanan, ``On approximating optimum repairs for
  functional dependency violations,'' in {\em ICDT}, pp.~53--62, 2009.

\bibitem{lopatenko2007efficient}
A.~Lopatenko and L.~Bravo, ``Efficient approximation algorithms for repairing
  inconsistent databases,'' in {\em ICDE}, pp.~216--225, 2007.

\bibitem{krishnan2016activeclean}
S.~Krishnan, J.~Wang, E.~Wu, M.~J. Franklin, and K.~Goldberg, ``Activeclean:
  Interactive data cleaning for statistical modeling,'' {\em {PVLDB}}, vol.~9,
  no.~12, pp.~948--959, 2016.

\bibitem{mayfield2010eracer}
C.~Mayfield, J.~Neville, and S.~Prabhakar, ``{ERACER}: {A} database approach
  for statistical inference and data cleaning,'' in {\em SIGMOD}, pp.~75--86,
  2010.

\bibitem{yakout2013don}
M.~Yakout, L.~Berti{-}{\'{E}}quille, and A.~K. Elmagarmid, ``Don't be scared:
  {U}se scalable automatic repairing with maximal likelihood and bounded
  changes,'' in {\em SIGMOD}, pp.~553--564, 2013.

\bibitem{prokoshynaSCMS15}
N.~Prokoshyna, J.~Szlichta, F.~Chiang, R.~J. Miller, and D.~Srivastava,
  ``Combining quantitative and logical data cleaning,'' {\em {PVLDB}}, vol.~9,
  no.~4, pp.~300--311, 2015.

\bibitem{rekatsinas2017holoclean}
T.~Rekatsinas, X.~Chu, I.~F. Ilyas, and C.~R{\'{e}}, ``Holoclean: Holistic data
  repairs with probabilistic inference,'' {\em {PVLDB}}, vol.~10, no.~11,
  pp.~1190--1201, 2017.

\bibitem{giannakopoulou2017cleanm}
S.~Giannakopoulou, M.~Karpathiotakis, B.~Gaidioz, and A.~Ailamaki, ``Clean{M}:
  An optimizable query language for unified scale-out data cleaning,'' {\em
  {PVLDB}}, vol.~10, no.~11, pp.~1466--1477, 2017.

\bibitem{niu2012deepdive}
F.~Niu, C.~Zhang, C.~R{\'e}, and J.~W. Shavlik, ``Deep{D}ive: {W}eb-scale
  knowledge-base construction using statistical learning and inference,'' {\em
  VLDS}, vol.~12, pp.~25--28, 2012.

\bibitem{domingos2009markov}
P.~M. Domingos and D.~Lowd, {\em Markov Logic: An Interface Layer for
  Artificial Intelligence}.
\newblock Synthesis Lectures on Artificial Intelligence and Machine Learning,
  Morgan {\&} Claypool Publishers, 2009.

\bibitem{niu2011tuffy}
F.~Niu, C.~R{\'{e}}, A.~Doan, and J.~W. Shavlik, ``Tuffy: Scaling up
  statistical inference in markov logic networks using an {RDBMS},'' {\em
  {PVLDB}}, vol.~4, no.~6, pp.~373--384, 2011.

\bibitem{fan2012foundations}
W.~Fan and F.~Geerts, {\em Foundations of Data Quality Management}.
\newblock Synthesis Lectures on Data Management, Morgan {\&} Claypool
  Publishers, 2012.

\bibitem{redman1998impact}
T.~C. Redman, ``the impact of poor data quality on the typical enterprise,''
  {\em Commun. {ACM}}, vol.~41, no.~2, pp.~79--82, 1998.

\end{thebibliography}

%\begin{IEEEbiography}[{\includegraphics[width=0.96in,height=1.2in,clip,keepaspectratio]
%{gcc.eps}}]{Congcong Ge}
%received the BS degree in computer science from Zhejiang University of Technology, China, in 2016. He is currently working toward the PhD degree in the College of Computer Science, Zhejiang University, China. His research interests include machine learning interaction with data management technology.
%\end{IEEEbiography}

\balance
\end{document}